%% file: ms.tex
\documentclass[numberedappendix]{emulateapj}

\usepackage{graphicx}
\usepackage{dcolumn}
\usepackage{bm}
\usepackage{epsf}
\usepackage{verbatim}
\usepackage{amsmath}
\usepackage{amsfonts}
\usepackage{ifthen}
\usepackage{epsfig}
\usepackage{verbatim}
\usepackage{dcolumn}

\usepackage{epstopdf}
\usepackage{amsmath}

\def\myfig#1{./#1}

\newboolean{incApp}
\setboolean{incApp}{false}
\newcommand{\mycond}[2]{\ifthenelse{#1}{#2}{}}
\newcommand{\myNcond}[2]{\ifthenelse{#1}{}{#2}}

\newcommand{\mydelta}{{l}}
\newcommand{\lpara}{{l_{3}}}
\newcommand{\initlpara}{{\init{l}_{3}}}
\newcommand{\init}[1]{{\tilde{#1}}}

\newcommand{\mystreamline}{s}

\newcommand{\llambda}{\initial{\mydelta}_\mysteadyj}
\newcommand{\ilLambda}{\initial{L}}

\newcommand{\mygconst}[1]{{\breve{#1}}}
\newcommand{\pcoma}{\mbox{ ;}}

\newcommand{\initBv}{\tilde{B}_3}
\newcommand{\deltats}{\Delta t}
\newcommand{\myrho}{\varrho}

\newcommand{\mythird}{q}
\newcommand{\lthird}{\delta_\mythird}
\newcommand{\initlthird}{\init{\delta}_\mythird}
\newcommand{\myfour}{Q}
\newcommand{\lfour}{\delta_{\mbox{\tiny Q}}}

\newcommand{\vfour}{v_{\mbox{\tiny Q}}}
\newcommand{\mya}{\bm{a}}
\newcommand{\myH}{\init{H}}
\newcommand{\mysteadyj}{\kappa}
\newcommand{\mypathk}{\alpha}
\newcommand{\mylambda}{\mygconst{R}}
\newcommand{\myshort}{\mygconst{\mymu}}
\newcommand{\mymu}{\mu}
\newcommand{\myPT}{P_{\mbox{\tiny th}}}
\newcommand{\myPTi}{\init{P}_{\mbox{\tiny th}}}
\newcommand{\Pram}{P_{\mbox{\tiny ram}}}
\newcommand{\Prami}{\init{P}_{\mbox{\tiny ram}}}
\newcommand{\Rey}{Re}
\newcommand{\myepsilon }{\zeta}
\newcommand{\myPLS}{L_m^2} 
\newcommand{\angleBM}{\theta}

\newcommand{\REarth}{R_\oplus}
\newcommand{\Mychi}{\chi}
\newcommand{\Myichi}{\init{\chi}}

\newcommand{\MyPB}{P_{\tiny\mbox{B}}}
\newcommand{\MyPBi}{\init{P}_{\tiny\mbox{B}}}
\newcommand{\fixapj}[1]{#1}

\input{Definitions.tex}

\makeatletter
\renewcommand*\env@matrix[1][\arraystretch]{%
  \edef\arraystretch{#1}%
  \hskip -\arraycolsep
  \let\@ifnextchar\new@ifnextchar
  \array{*\c@MaxMatrixCols c}}
\makeatother

\begin{document}

\title{Magnetohydrodynamics using path or stream functions}
\shorttitle{MHD with path/stream functions}
\shortauthors{Naor \& Keshet}
\author{Yossi Naor}

\author{Uri Keshet}

\affil{\footnotesize{Physics Department, Ben-Gurion University of the Negev, P.O.Box 653, Be'er-Sheva 84105, Israel; naoryos@post.bgu.ac.il}}

\date{\today}

\begin{abstract}
Magnetization in highly conductive plasmas is ubiquitous to astronomical systems. Flows in such media
can be described by three path functions $\Lambda_\alpha$, or, for a steady flow, by two stream functions $\lambda_\kappa$ and an additional field such as the mass density $\rho$, velocity $v$, or travel time $\Delta t$. While typical analyses of a frozen magnetic field $\boldsymbol{B}$ are problem-specific and involve nonlocal gradients of the fluid element position $\boldsymbol{x}(t)$, we derive the general, local (in $\Lambda$ or $\lambda$ space) solution $\boldsymbol{B}=(\partial\boldsymbol{x}/\partial\Lambda_\alpha)_t \tilde{B}_\alpha \rho/\tilde{\rho}$, where Lagrangian constants denoted by a tilde are directly fixed at a boundary hypersurface $\tilde{H}$ on which $\boldsymbol{B}$ is known.
For a steady flow, $\tilde{\rho}\boldsymbol{B}/\rho=(\partial\boldsymbol{x}/\partial\lambda_\kappa)_{\Delta t}\tilde{B}_\kappa+\boldsymbol{v}\tilde{B}_3/\tilde{v}$; here the electric field $\boldsymbol{E}\sim(\tilde{B}_2\boldsymbol{\nabla}\lambda_1-\tilde{B}_1\boldsymbol{\nabla}\lambda_2)/\tilde{\rho}$ depends only on $\lambda_\kappa$ and the boundary conditions.
Illustrative special cases include compressible axisymmetric flows and incompressible flows around a sphere, showing that viscosity and compressibility enhance the magnetization and lead to thicker boundary layers. 
Our method is especially useful for directly computing electric fields, and for addressing upstream magnetic fields that vary in spacetime.
We thus estimate the electric fields above heliospheres and magnetospheres, compute the draping of magnetic substructure around a planetary body, and demonstrate the resulting inverse polarity reversal layer.
\fixapj{Our analysis can be immediately incorporated into existing hydrodynamic codes that are based on stream or path functions, in order to passively evolve the electromagnetic fields in a simulated flow.
Furthermore, in such a prescription, the electromagnetic fields are frozen onto the grid, so it may be developed into a fully magnetohydrodynamic (MHD), efficient simulation.
}
\end{abstract}


\maketitle

\section{Introduction}
\label{sec:Introduction}

Magnetic fields frozen in highly conductive plasmas are studied in diverse astronomical systems.
Examples include the solar corona, flares and wind \citep[\eg][]{Longcope2005,ZhangLow2005}, planetary magnetospheres and bow shocks \citep[\eg][]{Spreiter70, SpreiterStahara95,Zhang_etal2004,Corona-Romero2013}, the interstellar medium (ISM) \cite[\eg][]{PriceBate2008,Lietal2011,Padovanietal2014}, in particular where it meets the solar wind \citep[][]{Parker61, Aleksashovetal2000, Whang2010}, and the intergalactic medium (IGM) of galaxy groups and galaxy clusters \citep[][]{BernikovSemenov79, Kim_etal91, Vikhlinin_etal2001,Keshet10, Bruggen2013}.
For general reviews of magnetization in astronomical systems, see \citet{Widrow2002}, \citet{Vallee2011}.

In such systems, magnetization away from shocks and reconnection regions can be typically approximated using ideal magnetohydrodynamics (MHD), on a background ranging from a simple axisymmetric flow around a blunt object to complicated, turbulent motions.
The passive evolution of magnetic fields frozen in a given flow is thus important in space physics, astrophysics, applied mathematics, and computational physics \citep[\eg][]{Ranger97,Sekhar03,Sekhar_etal_05,Bennett2008}.
However, present derivations of the magnetic field evolution in a general flow are inherently nonlocal.

Formally, the ideal MHD equations can be solved to give \citep{Elsasser56}
\begin{equation}
\boldsymbol{B}=\frac{\rho}{\initial{\rho}}\left(\initial{\boldsymbol{B}}\cdot\initial{\grad}\right)\boldsymbol{x}\coma\label{eq:Elsasser_B}
\end{equation}
where $\boldsymbol{B}$ is the magnetic field, $\rho$ is the mass density, $\boldsymbol{x}$ is the position of the fluid element, and a tilde denotes (henceforth) a Lagrangian constant evaluated on a reference hypersurface $\myH$, on which $\bm{B}$ is known.
Thus, $\initial{\grad}$ is a non-local operator, acting in the vicinity of $\myH$, rather than around $\bm{x}$.
Therefore, to derive $\bm{B}$ from Equation~(\ref{eq:Elsasser_B}), one must essentially integrate first over the flow to compute the mapping of $\bm{x}$ on $\myH$.

Here we show that $\bm{B}$ and $\bm{E}$ can be locally and simply computed for a general flow, by working in the space spanned by the path functions or stream functions that describe the dynamics.
Such functions exist as long as particle diffusion can be neglected \citep[][]{Yih57}.
Stream functions describe steady flows, and are also known as the Euler or Clebsch potentials of the mass flux \citep{Euler1769,Clebsch1859}. Path functions describe time-dependent flows, and are sometimes referred to as material functions \citep{Van-RoesselHui1991,LohHui2000}.

Thus, the dynamics of time dependent flows can be fully described by three path functions, $\Lambda_{\mypathk=1,2,3}$ \citep[][in three spatial dimensions; used henceforth and denoted 3D]{Yih57}.
The dynamics of steady flows can be fully described by two stream functions, $\lambda_{\mysteadyj=1,2}$, which fix the mass flux $\bm{j}\equiv \rho\bm{v}$ \citep{Giese51}, and an additional field $\varphi$ such as $\rho$, the velocity $v$, or the travel time $\Delta t$.
(Here we ignore an additional thermal quantity, such as temperature or pressure, which is needed to fully describe the flow but is unnecessary for our purposes.)
Path (stream) functions are useful in picturing the flow, because their equivalue surfaces intersect at pathlines (streamlines), the trajectories of fluid elements in spacetime (in space).

For simplicity, we focus on non-relativistic flows, with negligible particle diffusion, in the ideal MHD limit.
Arbitrary flows are considered, including those involving discontinuity surfaces such as shocks.
Our assumptions, in particular ideal MHD, may break down across shocks and in reconnection regions, where kinetic plasma effects can be important.
In such cases, our analysis may be piecewise applied, for example, on both sides of the shock.

The solution we derive for passive magnetization holds for arbitrary initial magnetic field configurations.
It is analyzed for a general steady, axisymmetric flow, tested for an arbitrary time dependent, one-dimensional (1D) flow, and illustrated for basic flow patterns, in particular steady flows around blunt objects used to model astronomical systems.
This includes both novel solutions and generalizations of previously solved problems.
The results indicate that passive magnetization is in general enhanced by viscosity and compressibility effects, leading to stronger magnetization in front of moving objects, and thus to thicker boundary layers.

Analytic solutions to the nonmagnetized flow around an obstacle are available only for a handful of time-independent cases, such as incompressible, potential flows around simple objects \citep[\eg][]{LandauLifshitz59_FluidMechanics,Leal2007}, or the approximate, compressible flow in front of an axisymmetric body \citep[][]{KeshetNaor2014}.
Solutions to the induced magnetization are even more rare, notable examples including the steady, incompressible, potential flow around a sphere \citep{ChackoHassam97,Lyutikov06, DursiPfrommer08,Romanelli14} or around simplified surfaces representing for example bow shocks \citep{Corona-Romero2013} or the heliopause \citep{Roken2014}.
We are unaware of a previous magnetization solution even for the simple, Stokes (creeping) incompressible steady flow around a sphere.
Such solutions are derived as special cases of our general axisymmetric result.

Our primary goal is to show that the evolution of the electromagnetic field in a given system becomes trivial (transparent) if the time-dependent (steady) flow is parameterized in terms of path (stream) functions.
This enables us to quickly reproduce known analytic results, and to analytically solve the field evolution in simple flows and test problems of astrophysical importance.
Moreover, our method can be incorporated into existing numerical simulations that utilize a stream function or path function description; this is straightforward for weak electromagnetic fields, but appears feasible for full MHD, at least in two-dimensions (2D).

The paper is organized as follows.
In Section \ref{sec:Magprofile}, we derive the temporal evolution of the magnetic field, for both steady and time dependent flows.
Arbitrary axisymmetric flows are analyzed in Section \ref{subsec:B_axisSym}, focusing in particular on the axis of symmetry.
In Section \ref{sec:special_cases}, we present some basic flows and analyze their magnetization.
In Section \ref{sec:Astrophysical applications}, we apply the analysis to a few illustrative astronomical systems.
The results are summarized and discussed in Section~\ref{sec:Discussion}\fixapj{, where we also examine the integration of our results into simulations that are based on path or stream functions, and the prospects for developing a full MHD simulation based on this approach.}

\section{Evolving electromagnetic fields}
\label{sec:Magprofile}

We begin in Section~\ref{sec:Magprofileintro} by introducing the MHD equations, and switching to a Lagrangian perspective.
Next, we analyze the evolution of $\bm{B}$ for arbitrary steady (in Section~\ref{sec:Magprofile_steady_state}) and time-dependent (in Section~\ref{sec:Magprofile_unsteady_state}) flows. To demonstrate the consistency of these two frameworks, in Section~\ref{sec:time_dependent_Streamfunctions}, an arbitrary steady flow is analyzed using the time-dependent formalism.

\subsection{Two Viewpoints of the Relation $\boldsymbol{B}\propto\rho\boldsymbol{l}$}\label{sec:Magprofileintro}
Let $\boldsymbol{l}$ be an infinitesimal vector adjoining nearby fluid elements that lie on the same magnetic field line (henceforth: length element). The magnetic field can be shown to evolve in proportion to $\rho \bm{l}$ \citep[\eg][]{LandauLifshitz60_Electrodynamics}.
We derive this relation in an Eulerian picture in Section~\ref{subsec:eulerian_pic}, and in a Lagrangian framework in Section~\ref{subsec:Lagrangian_pic}.

\subsubsection{Eulerian picture}\label{subsec:eulerian_pic}
The ideal MHD equations,
\begin{eqnarray}
& \mbox{Gauss' law,}\quad \quad  & \grad\cdot\boldsymbol{B}=0\pcoma\label{eq:Gauss}\\
& \mbox{continuity,}\quad \quad & \frac{\partial\rho}{\partial t}+\grad\cdot\left(\rho\boldsymbol{v}\right)=0\pcoma\label{eq:continuity}\\
& \mbox{convection,}\quad \quad  & \frac{\partial\boldsymbol{B}}{\partial t }=\grad\times\left(\boldsymbol{v}\times\boldsymbol{B}\right)\pcoma\label{eq:convection}\\
& \mbox{and Ohm's law,}\quad  & \boldsymbol{E}=-\frac{\boldsymbol{v}}{c}\times\boldsymbol{B}\label{eq:E_field}\coma
\end{eqnarray}
can be combined \citep[\eg][]{LandauLifshitz60_Electrodynamics} to give the Helmholtz equation
\begin{equation}\label{eq:Helmholtz}
\frac{d}{dt}\left(\frac{\boldsymbol{B}}{\rho}\right)=\left(\frac{\boldsymbol{B}}{\rho}\cdot\grad\right)\boldsymbol{v}\fin
\end{equation}
Here, $\bm{E}$ is the electric field, and $c$ is the speed of light.

As $\boldsymbol{l}$ is  fixed to the flow, it satisfies the equation
\begin{equation}\label{eq:fluidline}
\frac{d\boldsymbol{l}}{dt}=\left(\boldsymbol{l}\cdot\grad\right)\boldsymbol{v}\fin
\end{equation}
Comparing Equations~(\ref{eq:Helmholtz}) and (\ref{eq:fluidline}) indicates that as $(\boldsymbol{B}/\rho)$ and $\bm{l}$ evolve, their ratio remains constant, such that
\begin{equation} \label{eq:generalB}
 \boldsymbol{B}=\initial{B}\frac{\rho\boldsymbol{l}}{\initial{\rho}\initial{l}} \fin
  \end{equation}
This equation holds for both steady and time-dependent flows, and even across surfaces of discontinuity.
Using Equation~(\ref{eq:generalB}) to evolve $\bm{B}$ from $\myH$ guarantees that the MHD equations are satisfied everywhere, assuming continuity and that Gauss' law holds on $\myH$; see appendix \ref{sec:app_Gauss}.

In our non-relativistic, ideal MHD limit, $\boldsymbol{E}$ is small, and may be estimated from Equation~(\ref{eq:E_field}) once $\bm{B}$ is known.

\subsubsection{Lagrangian picture}\label{subsec:Lagrangian_pic}
For the subsequent analysis, it is useful to visualize the magnetic field amplitude as proportional to the density of field lines, \ie inversely proportional to the distance $\psi$ between them. Therefore, $\boldsymbol{B}$ in the direction $\unit{l}$ is inversely proportional to the perpendicular (to $\unit{l}$) area element $\mathcal{A}$ spanned by two such distances, $\bm{\psi}_1$ and $\bm{\psi}_2$. Hence,
 \begin{equation}
\boldsymbol{B}=\initial{B}\frac{\initial{\mathcal{A}}}{\mathcal{A}}\hat{\boldsymbol{l}} \coma
 \end{equation}
where $\mathcal{A}=\hat{\boldsymbol{l}}\cdot(\boldsymbol{\psi_1}\times\boldsymbol{\psi_2})$ is assumed infinitesimally small.
As the mass $\rho \mathcal{A} l$ of the fluid element in the parallelogram spanned by $\{\bm{l},\bm{\psi}_1,\bm{\psi}_2\}$ is constant, $\tilde{\mathcal{A}}/\mathcal{A}={\rho l}/(\tilde{\rho}\tilde{l})$, and we recover Equation~(\ref{eq:generalB}).

\subsection{Steady Flow Magnetization}
\label{sec:Magprofile_steady_state}
\subsubsection{General analysis}

The flow can be fully determined by specifying two stream functions, $\lambda_{\mysteadyj}$ where $\mysteadyj\in\{1,2\}$ (henceforth), and an additional scalar field $\varphi$ \citep{Giese51}.
Two surfaces, defined by constant values of the stream functions, intersect along a streamline, so
\begin{align}\label{eq:vlambdaperp}
\bm{v}\cdot\grad\lambda_\mysteadyj=0\fin
\end{align}
This equation, along with the time-independent continuity equation
 \begin{equation}\label{eq:stationary_continuity}
 \grad\cdot\left(\rho\bm{v}\right)=0\coma
 \end{equation}
 are identically satisfied by the $\lambda$ gauge \citep{Yih57}
\begin{equation}\label{eq:streamfunctionsdefinition}
\frac{\rho\bm{v}}{\mygconst{\rho}\mygconst{v}}
= \left(\grad\lambda_1\times\grad\lambda_2\right)
= \left| \begin{array}{ccc}
\hat{\bm{x}}_1& \hat{\bm{x}}_2 & \hat{\bm{x}}_3 \\\\
\frac{\partial\lambda_1}{\partial x_1} &\frac{\partial\lambda_1}{\partial x_2}  &\frac{\partial\lambda_1}{\partial x_3}  \\\\
\frac{\partial\lambda_2}{\partial x_1} &\frac{\partial\lambda_2}{\partial x_2}  &\frac{\partial\lambda_2}{\partial x_3}  \end{array} \right|\coma
\end{equation}
where the determinant, as written without Lam$\acute{e}$ coefficients, is valid only in Cartesian coordinates.
We use stream functions (and later, path functions) as coordinates.
For convenience, we take them in units of length by introducing the overall normalizations denoted by breve (henceforth; \eg $\rho/\mygconst{\rho}$ is a dimensionless mass density).
Equation (\ref{eq:streamfunctionsdefinition}), which is known as the Clebsch representation for the mass flux \citep[\eg][]{Stern67}, indicates that $\rho\bm{v}$ cannot be chosen as the independent $\varphi$, and that swapping $\lambda_1$ and $\lambda_2$ would reverse the flow.

In order to parameterize the position of fluid elements in the volume spanned by the flow, we must introduce, in addition to $\lambda_1$ and $\lambda_2$, a third variable $\mythird\left(\bm{x}\right)$, not necessarily equal to $\varphi$.
The surfaces defined by $\lambda_1$, $\lambda_2$ and $q$ may never overlap; namely, the vectors
\begin{equation} \label{eq:A_basis}
\bm{A}_1=\grad \lambda_1 \, ; \quad
\bm{A}_2=\grad \lambda_2 \, ; \quad
\bm{A}_3=\left(\frac{v}{v_\mythird}\right)\grad \mythird
\end{equation}
cannot be coplanar anywhere in the flow (except at stagnation points, where the analysis breaks down; see below).
Here, $v_\mythird\equiv d\mythird/dt=\bm{v}\cdot\grad\mythird$ is the rate of change in $\mythird$ along the flow (in units of $\mythird/t$).
The normalization factor $(v/v_\mythird)$ is included to keep the $\{\bm{A}_j\}$ basis dimensionless.
The freedom in the choice of $\mythird$ will be utilized in Section~\ref{subsubsec:GeneralChoices}.

It is useful to introduce the reciprocal basis $\mya_j=\bm{A}_j^{-1}$, defined \citep[\eg][]{Kishan2007} by $\mya_i \cdot \bm{A}_j=\delta_{ij}$, where $\delta_{ij}$ is Kronecker's delta.
Namely,
\begin{equation} \label{eq:a_basis}
\mya_1\equiv\left(\frac{\partial\bm{x}}{\partial\lambda_1}\right)_{\mythird} \pcoma \,\, \mya_2\equiv\left(\frac{\partial\bm{x}}{\partial\lambda_2}\right)_{\mythird} \pcoma  \,\,
\mya_3\equiv \frac{v_\mythird}{v}\left(\frac{\partial\bm{x}}{\partial\mythird}\right)_{\vec{\lambda}}=\hat{\bm{v}} \fin
\end{equation}
Here, $\vec{\lambda}\equiv\{\lambda_1, \lambda_2\}$, and it is understood (henceforth) that  derivatives with respect to $\lambda_1$ $(\lambda_2)$ are taken at a fixed $\lambda_2$ $(\lambda_1)$.
Like $\{\bm{A}_j\}$, the reciprocal basis too is dimensionless, and is nowhere coplanar.
Note that the $\mya_j$ are not necessarily perpendicular to one another, and that the $\mythird$-dependent $\mya_{1,2}$ are not necessarily unit vectors.

An advantage of the $\{\mya_j\}$ basis is that the $\mysteadyj=\{1,2\}$ components of any vector $\bm{V}$ in this basis are independent of $\mythird$, as
\begin{equation} \label{eq:GeneralProj}
\bm{V}= V_\mysteadyj \mya_\mysteadyj + V_3 \mya_3
= \begin{bmatrix}[1.3] V_1 \\ V_2 \\ V_3 \end{bmatrix}
= \begin{bmatrix}[1.3] \bm{V}\cdot \bm{A}_1\\ \bm{V}\cdot\bm{A}_2 \\ \bm{V}\cdot \bm{A}_3 \end{bmatrix}
= \begin{bmatrix}[1.3] \bm{V}\cdot\grad\lambda_1 \\ \bm{V}\cdot\grad\lambda_2 \\ \frac{v}{v_\mythird}\bm{V}\cdot \grad \mythird \end{bmatrix} \fin
\end{equation}
Moreover, for $\bm{l}$, the $\mysteadyj$ components $l_\mysteadyj$ are conserved along streamlines, as we show below.
We use (henceforth) square brackets to denote vectors in the reciprocal basis, which in general is neither orthogonal nor normalized, and reserve round brackets for vectors in orthonormal (in particular, cylindrical, except in Sections~\ref{sec:rarefaction_wave_Eample} and \ref{sec:spacetime_Btilde}) bases.

 Now, consider two infinitesimally close fluid elements, which at some instant are at the locations $\bm{x}(\lambda_\mysteadyj,\mythird)$ and
\begin{align}
\bm{x}+\bm{l}&=\bm{x}\left(\lambda_\mysteadyj+l_\mysteadyj,\mythird+\lthird\right)\\\nonumber
&=\bm{x}\left(\lambda_\mysteadyj,\mythird\right)+l_\mysteadyj\left(\frac{\partial \bm{x}}{\partial\lambda_\mysteadyj}\right)_{\mythird}+\lthird\left(\frac{\partial \bm{x}}{\partial\mythird}\right)_{\vec{\lambda}}+O\left(l^2\right)\coma
\end{align}
where we generalized the Einstein summation rule also for stream function indices $\mysteadyj=\{1,2\}$
(and, later on, for path function indices $\mypathk=\{1,2,3\}$).
As the $\vec{\lambda}$ of each of the two fluid elements is conserved along the flow, the intervals $l_\mysteadyj=\llambda$ are constant along the streamline.

\begin{figure}[t!]
\centering
\includegraphics[width=0.48\textwidth]{\myfig{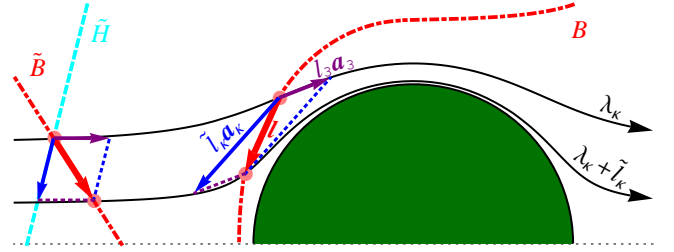}}
\caption{
Illustration of magnetic field evolution in the flow around a blunt object (green half disk), starting from a boundary hypersurface $\myH$ (dashed cyan curve).
A magnetic field line $\bm{B}$ (dotted-dashed red curve) is locally represented by an infinitesimal vector $\bm{l}$ (thick red arrow), adjoining two fluid elements (pink disks) that are initially along $\bm{B}$, and advected along streamlines (thin black arrows with stream function labels).
In the reciprocal basis, $\bm{l}$ is decomposed (see labels) into components parallel to the streamline ($l_3\mya_3$, purple arrows) and along the (constant $\mythird$) stream function gradients ($l_\mysteadyj \mya_\mysteadyj$, blue arrows; here initially confined to $\myH$ for the specific choice $\mythird=\Delta t$; see Section~\ref{subsubsec:GeneralChoices}).}
\label{fig:lambda_hat}
\vspace{2mm}
\end{figure}

Henceforth, we work to first order in the infinitesimal $l$, such that
\begin{equation}\label{eq:lsteady_general}
\bm{l}
=l_\mysteadyj \mya_\mysteadyj+\lpara\mya_3
=\llambda\mya_\mysteadyj+\lpara\mya_3
= \begin{bmatrix}[1.3] \init{l}_1 \\ \init{l}_2 \\ \lpara \end{bmatrix}
\coma
 \end{equation}
where $\lpara=v\lthird/v_\mythird$.
As $l_\mysteadyj=\llambda$, the $\mysteadyj$ components of $\bm{l}$ in the reciprocal basis are indeed seen to be constant along the streamline.
In contrast, $\lpara$ is in general not conserved.
The evolution of $\bm{l}$ and its decomposition in the reciprocal basis are illustrated in Figure~\ref{fig:lambda_hat}.

In order to evaluate $\lpara$, consider the travel time $\Delta t$ of a fluid element from $\myH$ to the position $\bm{x}$,
  \begin{equation}\label{eq:time_t}
  \Delta t\left(\bm{x}\right)=\int\limits_{\init{\mythird}(\bm{x})}^{\mythird\left(\bm{x}\right)}\frac{d\mythird'}{v_\mythird}\fin
  \end{equation}
During the same time, the $\bm{x}+\bm{l}$ fluid element travels from $\init{\mythird}+\initlthird$ (not necessarily on $\myH$) to $\mythird+\lthird$, so
 \begin{equation}\label{eq:DeltaTq}
  \int\limits_{\init{\mythird}}^{\mythird}\frac{d\mythird'}{v_\mythird\left(\lambda_\mysteadyj,\mythird'\right)}
  = \Delta t
  =\int\limits_{\init{\mythird}+\initlthird}^{\mythird+\lthird}
  \frac{d\mythird'}{v_\mythird\left(\lambda_\mysteadyj+\llambda,\mythird'\right)}\fin
 \end{equation}
 Expanding to first order in $l$ indicates that
   \begin{equation}\label{eq:lq}
  \frac{\lthird}{v_\mythird}=\frac{\initlthird}{\init{v}_\mythird}-\llambda \left( \frac{\partial\deltats}{\partial\lambda_\mysteadyj} \right)_{\mythird,\init{\mythird}} \coma
    \end{equation}
where derivatives of $\Delta t=\Delta t(\vec{\lambda}, \mythird, \init{\mythird})$ with respect to $\lambda_1$ or $\lambda_2$ are taken (henceforth) with the other three arguments of $\Delta t$ fixed.
The component $\lpara$ can now be related to the Lagrangian constants through
    \begin{equation}\label{eq:lv}
     \lpara=v\frac{\lthird}{v_q} = \initlpara\frac{v}{\init{v}}-v\llambda\frac{\partial\deltats}{\partial\lambda_\mysteadyj}\fin
        \end{equation}
If a surface of constant $\mythird$ overlaps with one of the constant $\lambda_\mysteadyj$ surfaces, $v_\mythird$ vanishes and Equation~(\ref{eq:lq}) is rendered invalid; a piecewise analysis may still be possible. The analysis breaks down at stagnation points, where $\Delta t$ (and so also $\bm{l}$ and $\bm{B}$) diverges.

Finally, the length element $\bm{l}$ is now given, throughout the flow connected to $\myH$, by
  \begin{equation}\label{eq:vecl_Deltaq}
   \bm{l}=\llambda\mya_\mysteadyj+
   \frac{\bm{v}}{\init{v}}\left(\initlpara-\llambda\init{v}\frac{\partial\deltats}{\partial\lambda_\mysteadyj}\right)
   = \begin{bmatrix}[1.3] \init{l}_1 \\ \init{l}_2 \\
   \frac{v}{\init{v}}\initlpara \end{bmatrix}
-  \begin{bmatrix}[1.3] 0 \\ 0 \\ v  \end{bmatrix} \llambda\frac{\partial\deltats}{\partial\lambda_\mysteadyj} \fin
  \end{equation}
Note that in addition to $\mya_\mysteadyj$, the $\Delta t$ derivative also depends on $\mythird$. It can be written as
\begin{equation}\label{eq:General_Deltat}
\frac{\partial\Delta t}{\partial\lambda_\mysteadyj} \equiv \left(\frac{\partial\Delta t}{\partial\lambda_\mysteadyj}\right)_{q,\init{q}}
=\int\limits_{\init{q}}^{q}\left(\frac{\partial v_q^{-1}}{\partial\lambda_\mysteadyj}\right)_{q'}dq'\fin
\end{equation}

The Lagrangian constants are fixed at $\tilde{\bm{x}}$, where the streamline meets $\myH$.
The point $\tilde{\bm{x}}(\bm{x})$ can be computed by inverting $\myH(\tilde{\bm{x}})$, as $\vec{\lambda}(\bm{x})=\vec{\lambda}(\init{\bm{x}})$ is given, and remains constant along the streamline.
The components $\init{l}_j$ are found by projecting $\init{\bm{l}}$ onto the reciprocal basis as in Equation~(\ref{eq:GeneralProj}),
\begin{equation} \label{eq:linitlambdakj}
\init{l}_\mysteadyj=\bm{l}\cdot \grad\lambda_\mysteadyj=\left(\bm{l}\cdot \grad\lambda_\mysteadyj\right)_\myH \quad \mbox{and} \quad
\init{l}_3=\left(\frac{v}{v_\mythird}\,\bm{l}\cdot \grad{q}\right)_\myH  \fin
\end{equation}

Equations~(\ref{eq:generalB}) and (\ref{eq:vecl_Deltaq}) yield the magnetic field, as
\begin{equation}\label{eq:vecB_Deltaq}
\frac{\initial{\rho}}{\rho} \bm{B}=\initial{B}\frac{\boldsymbol{l}}{\initial{l}}
    = \begin{bmatrix}[1.3] \init{B}_1 \\ \init{B}_2 \\   \frac{v}{\init{v}}\init{B}_3 \end{bmatrix}
    -  \begin{bmatrix}[1.3] 0 \\ 0 \\ v  \end{bmatrix}
    \tilde{B}_\mysteadyj \frac{\partial\deltats}{\partial\lambda_\mysteadyj}
    \coma
\end{equation}
where the Lagrangian constants $\init{B}_j=(\init{B}/\init{l})\init{l}_j$ are fixed by the projection Equation~(\ref{eq:GeneralProj}) on $\myH$,
\begin{equation} \label{eq:Binitlambdakj}
\init{B}_\mysteadyj=\left(\bm{B}\cdot \grad\lambda_\mysteadyj\right)_\myH \quad \mbox{and} \quad
\init{B}_3=\left(\frac{v}{v_\mythird}\,\bm{B}\cdot \grad{q}\right)_\myH  \fin
\end{equation}

Next, consider the electric field $\bm{E}$. The vector product $(\bm{v}\times\bm{B})$ in Equation~(\ref{eq:E_field}) eliminates the $\mya_3$ terms in Equation~(\ref{eq:vecB_Deltaq}), leaving
 \begin{align}\label{eq:ESteady}
 \bm{E}=\frac{\mygconst{\rho}\mygconst{v}}{\init{\rho}c}\left(\init{B}_2\grad\lambda_1-\init{B}_1\grad\lambda_2\right)\fin
 \end{align}
Therefore, $\bm{E}$ is entirely determined by the stream functions, independent of $\varphi$, once $\bm{B}$ and $\rho$ are given on $\myH$.
Notice that deriving the electric field directly from the stream (and below, from the path) functions is computationally much simpler than first computing the magnetic field, which typically involves integration, and then using Ohm's law to find $\bm{E}$.

\subsubsection{Choice of a third variable $\mythird(\bm{x})$}
\label{subsubsec:GeneralChoices}

 There are various choices for the third variable, $\mythird$, depending on the specific problem at hand.
A useful choice is the travel time $\Delta t$, which in some cases is easily estimated or computed. For $\mythird=\Delta t$, which is the specific case illustrated in Figure~\ref{fig:lambda_hat}, Equations~(\ref{eq:vecl_Deltaq}) and~(\ref{eq:vecB_Deltaq}) reduce to
\begin{equation}\label{eq:vecl_DeltaT}
\bm{l}= \llambda\left(\frac{\partial\bm{x}}{\partial\lambda_\mysteadyj}\right)_{\Delta t} + \initlpara\frac{\boldsymbol{v}}{\initial{v}}
= \begin{bmatrix}[1.3] \init{l}_1 \\ \init{l}_2 \\
\frac{v}{\init{v}}\initlpara \end{bmatrix}
\coma
\end{equation}
and
\begin{equation}\label{eq:vecB_DeltaaT}
\bm{B}= \left[\init{B}_\mysteadyj\left(\frac{\partial\bm{x}}{\partial\lambda_\mysteadyj}\right)_{\Delta t}+ \initBv\frac{\boldsymbol{v}}{\initial{v}}\right] \frac{\rho}{\init{\rho}}
=
\begin{bmatrix}[1.3] \init{B}_1 \\ \init{B}_2 \\   \frac{v}{\init{v}}\init{B}_3 \end{bmatrix}
\frac{\rho}{\initial{\rho}}
\fin
\end{equation}

Notice that Equation~(\ref{eq:vecl_DeltaT}) could have been directly inferred from the reciprocal basis projection Equation~(\ref{eq:GeneralProj}), by noting that $\bm{l}\cdot\grad(\Delta t)=\Delta t(\bm{x}+\bm{l})-\Delta t(\bm{x})+O(l^2)$ is constant (to first order) along a stream line, as this is the difference between the travel times of two nearby fluid elements.

In the case of an irrotational flow, a potential $\Phi$ can be defined such that $\bm{v}=\mygconst{v}\grad\Phi$, where the normalization factor assigns $\Phi$ with units of length.
Here we may choose $\mythird=\Phi$, and read $\bm{l}$ and $\bm{B}$ off Equations~(\ref{eq:vecl_Deltaq}) and (\ref{eq:vecB_Deltaq}), respectively, with the time derivative term (\cf Equation~(\ref{eq:General_Deltat}))
\begin{equation}
\frac{\partial\Delta t}{\partial\lambda_\mysteadyj}
=
\left(\frac{\partial\Delta t}{\partial\lambda_\mysteadyj}\right)_{\Phi,\init{\Phi}}
=
\mygconst{v}\int\limits_{\init{\Phi}}^{\Phi}{d\Phi'}\left(\frac{\partial v^{-2}}{\partial \lambda_\mysteadyj}\right)_{\Phi'}\fin
\end{equation}
Equipotential surfaces are perpendicular to the streamlines, so here $\mya_1$ and $\mya_2$ are conveniently perpendicular to $\mya_3$.
In simple, \eg axisymmetric, flows where $\mya_1$ and $\mya_2$ are perpendicular to each other, the reciprocal basis thus becomes orthogonal.

Among the infinite other possible choices of $\mythird$, noteworthy are the natural coordinates of the flow, for example the axial coordinate $z$ in simple axisymmetric flows, and the length $\mystreamline(\bm{x})$ of the streamline from $\myH$ to $\bm{x}$. In general, the simplest variable which is monotonic along the flow is advantageous.

\subsection{Time-dependent Flow Magnetization}
\label{sec:Magprofile_unsteady_state}

A time-dependent flow can be fully specified by three path functions, $\Lambda_{\mypathk=1,2,3}$, when particle diffusion can be neglected \citep{Yih57}.
Three spacetime manifolds, defined by some constant values of these path functions, intersect along a pathline.
Therefore, along the flow
\begin{equation}\label{eq:dLambdadt0}
\frac{d\Lambda_\mypathk\left(\bm{x},t\right)}{dt}=\left(\frac{\partial }{\partial t}+\bm{v}\cdot\grad\right)\Lambda_\mypathk\left(\bm{x},t\right)=0\coma
\end{equation}
which determines $\bm{v}$ through
\begin{equation}\label{eq:pathv}
\bm{v}=-\frac{\epsilon_{ijk}\left(\partial_t\Lambda_i\right) \left(\grad\Lambda_j\times\grad\Lambda_k\right)}{2\left(\grad\Lambda_1\times\grad\Lambda_2\right)\cdot\grad\Lambda_3}\coma
\end{equation}
where $\epsilon_{ijk}$ is the Levi-Civita symbol.
The continuity equation (\ref{eq:continuity}) is then identically satisfied by the $\Lambda$ gauge \citep{Yih57}
\begin{equation}\label{eq:pathrho}
\frac{\rho}{\mygconst{\rho}}=-\left(\grad\Lambda_1\times\grad\Lambda_2\right)\cdot\grad\Lambda_3\fin
\end{equation}
Equations~(\ref{eq:pathv}) and (\ref{eq:pathrho}) can be written as \citep{Yih57}
\begin{equation}\label{eq:pathfunctionsdefinition}
\frac{\rho}{\mygconst{\rho}} \begin{pmatrix} 1 \\ \bm{v} \end{pmatrix}
=\frac{\rho}{\mygconst{\rho}}\left(\bm{1}+\bm{v}\right)
=\left| \begin{array}{cccc}
\hat{\bm{x}}_1& \hat{\bm{x}}_2 & \hat{\bm{x}}_3  & \bm{1} \\\\
\frac{\partial\Lambda_1}{\partial x_1} &\frac{\partial\Lambda_1}{\partial x_2}  &\frac{\partial\Lambda_1}{\partial x_3} & \frac{\partial\Lambda_1}{\partial t} \\\\
\frac{\partial\Lambda_2}{\partial x_1} &\frac{\partial\Lambda_2}{\partial x_2}  &\frac{\partial\Lambda_2}{\partial x_3} & \frac{\partial\Lambda_2}{\partial t}\\\\
\frac{\partial\Lambda_3}{\partial x_1} &\frac{\partial\Lambda_3}{\partial x_2}  &\frac{\partial\Lambda_3}{\partial x_3} & \frac{\partial\Lambda_3}{\partial t}\end{array} \right|
\coma
\end{equation}
where $\bm{1}$ is a unit vector in some non-spatial dimension, and the determinant, as written without Lam$\acute{e}$ coefficients, holds only in Cartesian coordinates.
Equivalently \citep{Van-RoesselHui1991}, if $v^\mu\equiv(1,\bm{v})$,
\begin{equation}\label{eq:pathfunctionsdefinitionB}
\frac{\rho}{\mygconst{\rho}}\,v^\alpha
=-\epsilon^{\alpha\beta\gamma\delta}\frac{\partial\Lambda_1\partial\Lambda_2\partial\Lambda_3}{\partial x^\beta\partial x^\gamma\partial x^\delta} \coma
\end{equation}
where $\{\alpha,\beta,\gamma,\delta\}$ are four-indices, $x^0\equiv t$, and $\epsilon^{\alpha\beta\gamma\delta}$ is the four-dimensional Levi-Civita symbol.

To parameterize spacetime using the path functions $\Lambda_1$, $\Lambda_2$, and $\Lambda_3$, we normalize them in units of length, and introduce a fourth coordinate, $\myfour(\bm{x},t)$.
The fourth coordinate can be chosen arbitrarily, as long as the equivalue hypersurfaces of the four parameters never overlap.

Define $\bm{l}$ as the distance between two simultaneous, infinitesimally close spacetime events,
\begin{equation}
x^{\mu}\left(\Lambda_\mypathk,\myfour\right)=\left( \begin{array}{cc}
t\\
 \bm{x}\end{array} \right)
\end{equation}
 and
 \begin{align}
\begin{pmatrix} t \\ \bm{x}+\bm{l} \end{pmatrix} = x^{\mu}+dx^{\mu}&=x^{\mu}\left(\Lambda_\mypathk+L_\mypathk,\myfour+\lfour\right)\\\nonumber
&=x^{\mu}+L_\mypathk\frac{\partial x^{\mu} }{\partial\Lambda_\mypathk} +\lfour\frac{\partial x^{\mu} }{\partial\myfour}
+O(l^2) \fin
 \end{align}
As $x^\mu=x^\mu(\Lambda_\mypathk,\myfour)$, derivatives with respect to $\myfour$ or one of the $\Lambda_\mypathk$ are understood as taken with the other three arguments fixed.
Here, the infinitesimal $L_\mypathk=\ilLambda_\mypathk$ are constant along the pathline, because they are the differences between the conserved path function values.
In contrast, $\lfour$ varies in general along the pathline.

Working to first order in $l$, the spacetime length element is now given by
 \begin{align}
\left( \begin{array}{cc}
  0\\
   \bm{l}\end{array} \right)&=\ilLambda_\mypathk\left( \begin{array}{cc}
   \partial t/\partial\Lambda_\mypathk\\
    \partial \bm{x}/\partial\Lambda_\mypathk\end{array} \right)+\frac{\lfour}{\vfour}\left(\begin{array}{cc}
     1\\
     \bm{v}\end{array} \right)\coma
 \end{align}
where we defined $\vfour\equiv d\myfour/dt$ as the rate of change in $\myfour$ along the flow (in units of $\myfour/t$).
The vanishing temporal component requires
\begin{equation}
\frac{\lfour}{\vfour}= -\ilLambda_\mypathk\frac{\partial t}{\partial\Lambda_\mypathk}\coma
\end{equation}
so the spatial component gives
\begin{align}\label{eq:l_pathfour}
\bm{l}=\ilLambda_\mypathk\left(\frac{\partial \bm{x}}{\partial\Lambda_\mypathk}-\bm{v}\frac{\partial t}{\partial\Lambda_\mypathk}\right)\fin
\end{align}
Taking the scalar product of this result with $\grad \Lambda_\mypathk$, and using Equation~(\ref{eq:dLambdadt0}) and the chain rule, implies that the constant $\ilLambda_\mypathk$ satisfy
\begin{equation}\label{eq:initlpath}
\ilLambda_\mypathk=\bm{l}\cdot\grad\Lambda_\mypathk = \left(\bm{l}\cdot\grad\Lambda_\mypathk \right)_\myH \coma
\end{equation}
and are independent of $\myfour$.
They may thus be determined at any position where $\bm{l}$ is known, in particular on $\myH$.

Finally, combining Equations~(\ref{eq:generalB}) and (\ref{eq:l_pathfour}) yields
\begin{align}\label{eq:B_pathfour}
\bm{B}= \init{B}_\mypathk\frac{\rho}{\init{\rho}}\left(\frac{\partial \bm{x}}{\partial\Lambda_\mypathk}-\bm{v}\frac{\partial t}{\partial\Lambda_\mypathk}\right)\coma
\end{align}
where the Lagrangian constants $\init{B}_\mypathk\equiv(\init{B}/\init{l})\ilLambda_\mypathk$ can be determined using
\begin{equation}\label{eq:initBpath}
\init{B}_\mypathk
=\frac{\init{\rho}}{\rho}\bm{B}\cdot\grad\Lambda_\mypathk
=\left(\bm{B}\cdot\grad\Lambda_\mypathk \right)_\myH \fin
\end{equation}

As in the time-independent case, here too there is considerable freedom in choosing $\myfour$.
A useful choice is the temporal coordinate, $\myfour=t$, for which Equation~(\ref{eq:l_pathfour}) reduces to
\begin{align}\label{eq:l_patht}
\bm{l}=\ilLambda_\mypathk\left(\frac{\partial \bm{x}}{\partial\Lambda_\mypathk}\right)_t\fin
\end{align}
The time-dependent vectors $({\partial \bm{x}}/{\partial\Lambda_\mypathk})_t$ can now be identified as the generalizations of the reciprocal basis vectors found for steady flows; \cf Equations~(\ref{eq:a_basis}) and (\ref{eq:lsteady_general}).
Equation~(\ref{eq:B_pathfour}) now gives the magnetic field,
\begin{align}\label{eq:B_patht}
\bm{B}= \init{B}_\mypathk\frac{\rho}{\init{\rho}}\left(\frac{\partial \bm{x}}{\partial\Lambda_\mypathk}\right)_t\fin
\end{align}
Other choices of $\myfour$ may be advantageous in certain circumstances.

The electric field in a time-dependent flow becomes, according to Equations~(\ref{eq:E_field}) and (\ref{eq:B_pathfour}),
\begin{align}\label{eq:Etimedependent}
\bm{E}=\epsilon_{ijk}\frac{\mygconst{\rho}\init{B}_i}{\init{\rho}c}\left(\frac{\partial\Lambda_j}{\partial t}\right)\grad\Lambda_k\fin
\end{align}

\subsection{Time-dependent Description of a Steady Flow}\label{sec:time_dependent_Streamfunctions}
To ascertain that Section~\ref{sec:Magprofile_steady_state} and Section~\ref{sec:Magprofile_unsteady_state} are mutually consistent, we consider an arbitrary steady flow, and regard it as time-dependent.
If the steady flow is characterized by the stream functions $\lambda_1$ and $\lambda_2$, we can choose two path functions as time independent,
\begin{align}
\Lambda_1&=\lambda_1\quad \mbox{and} \quad \Lambda_2=\lambda_2\fin
\end{align}
Solving  Equations~(\ref{eq:pathv}) and (\ref{eq:pathrho}) with the aid of Equation~(\ref{eq:streamfunctionsdefinition}) and the definitions of Section~\ref{sec:Magprofile_steady_state}, we find that the third path function must be
\begin{align}
\Lambda_3&=\mygconst{v}\left(t-\deltats\right)\fin
\end{align}
We can now use the time-dependent results of Section~\ref{sec:Magprofile_unsteady_state}, and attempt to reproduce the steady flow results of Section~\ref{sec:Magprofile_steady_state}.

For simplicity, we choose $Q=t$, such that the solution Equation~(\ref{eq:l_patht}) for $\bm{l}$ in the time dependent regime gives
\begin{align}\label{eq:l_vec1_t}
\bm{l}
&=\ilLambda_\mysteadyj\left(\frac{\partial\bm{x}}{\partial\Lambda_\mysteadyj}\right)_{\Lambda_3,t}+ \ilLambda_3\left(\frac{\partial\bm{x}}{\partial\Lambda_3 }\right)_{\Lambda_\mysteadyj,t}\\\nonumber
&=\ilLambda_\mysteadyj\left(\frac{\partial\bm{x}}{\partial\lambda_\mysteadyj}\right)_{\deltats,t}+ \left(-\ilLambda_3 \frac{\init{v}}{\mygconst{v}}\right)\frac{\bm{v}}{\init{v}}
\fin
\end{align}
This is formally identical to Equation~(\ref{eq:vecl_DeltaT}), derived for a steady flow under the choice $q=\Delta t$, provided that one may consistently identify $\init{l}_\mysteadyj=\init{L}_\mysteadyj$ and $\init{l}_3=-\ilLambda_3( {\init{v}}/{\mygconst{v}})$.
Indeed, Equation~(\ref{eq:initlpath}) yields $\init{L}_\mysteadyj=\bm{l}\cdot\grad \Lambda_\mysteadyj=\bm{l}\cdot\grad \lambda_\mysteadyj=\init{l}_\mysteadyj$, and, on $\myH$,
$\init{L}_3=\bm{l}\cdot\grad \Lambda_3=-\mygconst{v}\bm{l}\cdot\grad(\Delta t) = -({\mygconst{v}}/{\init{v}})(v/v_q)\bm{l}\cdot\grad q=-({\mygconst{v}}/{\init{v}})\tilde{l}_3$; consistent with the definitions of Equation~(\ref{eq:linitlambdakj}).

In conclusion, the time-dependent analysis with $Q=t$ yields the time-independent analysis with $q=\Delta t$, when applied to a steady flow.
This shows that the two formalisms, involving path functions and stream functions, are consistent with each other.

With the above $\Lambda_\mypathk$, the time-dependent electric field in Equation~(\ref{eq:Etimedependent}) directly reduces to the time-independent $\bm{E}$ in  Equation~(\ref{eq:ESteady}), as $\partial\Lambda_j/\partial t=\mygconst{v} \delta_{3,j}$.

 \section{Steady Axisymmetric flow Magnetization}
 \label{subsec:B_axisSym}

  \subsection{General Analysis in Cylindrical Coordinates  }\label{sec:Spherical coordinate analysis}

Consider a steady, axisymmetric flow.
Using cylindrical coordinates $\left\{\myrho,\phi,z\right\}$, such flows are characterized by $v_\phi=0$, and by vanishing $\phi$ derivatives of the flow parameters. Therefore, constant $\phi$ surfaces are perpendicular to the flow, and we may choose $\lambda_1\equiv\mylambda\phi$, where $\mygconst{R}$ is some arbitrary length scale of the problem.
Since the velocity field $\bm{v}$ is $\phi$-independent, we may choose $\lambda_2\equiv\lambda(\myrho,z)$.
For these stream functions,  Equation~(\ref{eq:streamfunctionsdefinition}) yields
\begin{align}
\bm{v}= \left(v_\myrho,0,v_z\right)&=\frac{\mygconst{\mymu}}{\myrho\rho}\left(\frac{\partial\lambda}{\partial z},0,-\frac{\partial\lambda}{\partial \myrho}\right)\label{eq:lambdas_v}\coma
\end{align}
where $\bm{\mymu}\equiv \rho\myrho \bm{v}$ is the mass flux in a ring of radius $\myrho$ (and has units of viscosity).
Here, the normalizations $\mylambda$ and $\mygconst{\mymu}\equiv\mygconst{\rho}\,\mygconst{v}\,\mylambda$ arise from our use of stream functions as coordinates with units of length.
Round bracket vectors in this section pertain to cylindrical coordinates.

In addition to $\lambda_1$ and $\lambda_2$,  we must now choose a third variable $q$ to parameterize space.
The choice $\mythird=z$ is advantageous for simple flow regions in which $v_z$ does not change sign.
For our choice of stream functions and $q$, the reciprocal basis Equation~(\ref{eq:a_basis}) becomes
\begin{align}\label{eq:axisymmetric_a_i}
\mya_1=\frac{\myrho}{\mygconst{R}}\hat{\bm{\phi}}\pcoma\qquad \mya_2=-\frac{\myshort}{\mymu_z}\hat{\bm{\myrho}}\pcoma\qquad \mya_3=\hat{\bm{v}}\fin
\end{align}
A given $\init{\bm{B}}=(\init{B}_\myrho,\,\init{B}_\phi,\,\init{B}_z)$ may now be decomposed as $\init{B}_i\init{\mya}_i$, where (\cf Equation~(\ref{eq:Binitlambdakj}))
\begin{align}\label{eq:Binit_cylindrical}
\init{\bm{B}} = \begin{bmatrix}[1.3] \init{B}_1 \\ \init{B}_2 \\ \init{B}_3 \end{bmatrix}
= \begin{bmatrix}[1.3] \init{B}_\phi\frac{\mylambda}{\init{\myrho}} \\ ({\init{\mymu}_\myrho\init{B}_z-\init{\mymu}_z\init{B}_\myrho})/{\myshort} \\
\init{B}_z\frac{\init{v}}{\init{v}_z} \end{bmatrix}
\fin
\end{align}
Recall that in our notations, vectors in square brackets (as in Equation~(\ref{eq:Binit_cylindrical})) are in the reciprocal basis.

Equation (\ref{eq:vecB_Deltaq}) then yields
\begin{align}\label{eq:B_Cylindrical}
\frac{\init{\rho}}{\rho}\bm{B} & =
 \frac{\myrho}{\mylambda}\init{B}_1\hat{\bm{\phi}}-\frac{\mygconst{\mymu}}{\mymu_z}\init{B}_2\hat{\bm{\myrho}}
 +\frac{\bm{v}}{\init{v}}\left(\init{B}_3-\init{B}_2\init{v}\frac{\partial\deltats}{\partial\lambda}\right) \\
& = \begin{pmatrix}[1.7]
       \frac{\init{\mymu}_z}{\mymu_z} \init{B}_\myrho \\
       \frac{\myrho}{\init{\myrho}} \init{B}_\phi \\
       \frac{v_z}{\init{v}_z} \init{B}_z
  \end{pmatrix}
   +\begin{pmatrix}[1.7]
        \frac{v_\myrho}{\init{v}_z}-\frac{\init{\mymu}_\myrho}{\mymu_z} \\ 0 \\ 0
\end{pmatrix} \init{B}_z
- \begin{pmatrix}[1.7] v_\myrho \\ 0 \\ v_z \end{pmatrix} \init{B}_2 \frac{\partial\Delta t}{\partial\lambda}
\nonumber    \fin
\end{align}
The only computation necessary is an integral for the time derivative term,
\begin{equation}\label{eq:Deltat_cylindrical}
\frac{\partial\Delta t}{\partial\lambda} = \left[\frac{\partial\Delta t(\myrho,z)}{\partial\lambda}\right]_{\phi,z,\init{z}}=\int\limits_{\init{z}}^{z}\left(\frac{\partial v_z^{-1}}{\partial\lambda}\right)_{\phi,z'}dz'\fin
\end{equation}

The electric field is found from Equations~(\ref{eq:ESteady}), (\ref{eq:lambdas_v}) and (\ref{eq:Binit_cylindrical}), and given simply by
\begin{align}\label{eq:E_Cylindrical}
\bm{E} =\frac{\init{v}}{ c}\begin{pmatrix}[1.7]
       \frac{\mymu_z}{\init{\mymu}}\init{B}_\phi \\
       \frac{\init{\myrho}\mygconst{\mymu}}{\myrho\init{\mymu}}\init{B}_2 \\
      - \frac{\mymu_\myrho}{\init{\mymu}}\init{B}_\phi
  \end{pmatrix}
  =\frac{\init{v}}{\init{\mymu} c}\begin{pmatrix}[1.7]
       \mymu_z\init{B}_\phi \\
       \frac{\init{\myrho}}{\myrho}\, ({\init{\mymu}_\myrho\init{B}_z - \init{\mymu}_z \init{B}_\myrho}) \\
       -\mymu_\myrho\init{B}_\phi
  \end{pmatrix}
    \fin
\end{align}

\subsection{The Axis of Symmetry}
\label{subsec:B_along_the_axis}

Next, we focus on the vicinity of the axis of symmetry, $\myrho=0$, along which $v_\myrho=0$.
Taking the incident flow in the ($-\unit{z}$) direction, we denote the incoming velocity as $u(\rho\simeq 0,z)\equiv -v_z$, and assume it is finite.
Then, Equation~(\ref{eq:lambdas_v}) indicates that in the vicinity of the axis, $\lambda=(\rho u/2\myshort)\myrho^2 + O(\myrho^3)$.
Along a streamline, $\lambda$ is constant,therefore so is $\rho u \myrho^2$; this is a direct expression of mass conservation.
In the vicinity of the axis, we may thus write $\myrho/\init{\myrho}=(\init{\rho}\init{u}/\rho u)^{1/2}$ along a streamline.
This holds even in the presence of a shock, which conserves the normal, in this case axial, mass flux $\myj=\rho u$.

Equation ~(\ref{eq:B_Cylindrical}) now becomes
\begin{align} \label{eq:AxialB}
\bm{B}=
\begin{pmatrix}[1.7]
 \left(\frac{\rho/u}{\init{\rho}/\init{u}}\right)^{\frac{1}{2}} \init{B}_\myrho \\
 \left(\frac{\rho/u}{\init{\rho}/\init{u}}\right)^{\frac{1}{2}} \init{B}_\phi  \\
      \frac{\myj}{\init{\myj}} \init{B}_z
  \end{pmatrix}
  + C\begin{pmatrix}[1.7]
  0\\0\\ \init{B}_\myrho
  \end{pmatrix}
  \coma
\end{align}
where
\begin{equation} \label{eq:AxialC}
C(z)\equiv -\frac{\myj\init{\mymu}_z}{\init{\rho}\myshort} \frac{\partial\Delta t}{\partial\lambda}=
\frac{\myj\init{u}}{\sqrt{\init{\myj}}}\int\limits_{\init{z}}^{z}\frac{u_1dz'}{u_0^2\sqrt{\myj}} \coma
\end{equation}
and we expanded $u=u_0(z)+u_1(z)\rho+O(\rho^2)$.

For a potential flow, $u_1=0$, and so $C=0$.
Such a flow arises, for example, if a homogeneous incident flow remains subsonic or mildly supersonic \citep[\eg][]{LandauLifshitz59_FluidMechanics}.

It is typically possible to derive $(\rho/\init{\rho})$ as a function of $u$, such that $\bm{B}=\bm{B}(u)$ along the axis.
For example, consider a steady, inviscid flow in a polytropic gas of an adiabatic index $\gamma$, with dynamically insignificant magnetic field.
Such a flow is governed, in addition to continuity, by Euler's equation,
\begin{equation}\label{eq:Euler}
  \left(\boldsymbol{v}\cdot\grad\right)\boldsymbol{v}=-\frac{\grad P}{\rho}=-\frac{c_s^2\,\grad\rho}{\rho}\coma
\end{equation}
and Bernoulli's equation,
\begin{equation} \label{eq:Bernulli}
\frac{u^2}{2} + \frac{c_s^2}{\gamma-1} = \frac{\bar{c}_s^2}{\gamma-1} = \mbox{constant .}
\end{equation}
Here, $P$ is the fluid pressure, $c_s=(\gamma P/\rho)^{1/2}$  is the (local) speed of sound, and a bar denotes (henceforth) a putative stagnation point where $v=0$, whether or not such a point lies along a given streamline.

The axial mass density solution is $\rho^2\propto(S^2-M_0^2)^{S^2}$, where $S\equiv [2/(\gamma-1)]^{1/2}$, and $M_0\equiv u/\bar{c}_s$ is the axial Mach number with respect to the stagnation sound speed.
Hence, away from shocks,
\begin{equation} \label{eq:rho}
\frac{\rho}{\initial{\rho}}=\left(\frac{S^2-M_0^2}{S^2-\initial{M}_{0}^2}\right)^{1/(\gamma-1)}
\end{equation}
can be used in Equation~(\ref{eq:AxialB}) to find $\bm{B}$ once $u$ is determined.
When a shock is present, the Lagrangian constants here are taken downstream; if $\myH$ is located upstream, the shock compression ratio $(\init{M}_0/W)^2$ should be incorporated.
Here, $W\equiv [2/(\gamma+1)]^{1/2}$ and $S$ are the weak and strong shock limits of $M_0$; see \cite{KeshetNaor2014}.

\section{Magnetization of basic flows}
\label{sec:special_cases}

To illustrate the general magnetization solution, derived in general in Section~\ref{sec:Magprofile} and for axial symmetry in Section~\ref{subsec:B_axisSym}, here we study the magnetic fields evolving in specific, basic flows.
Passive magnetization in steady, incompressible flows around a sphere are analyzed in the irrotational (and thus, effectively inviscid) case in Section~\ref{subsec:Incompete}, and in the viscous limit in Section~\ref{subsec:Stokes}.
In Section~\ref{sec:rarefaction_wave_Eample} we study the simple, 1D class of time-dependent flows.

\subsection{Potential Incompressible Steady Flow Around a Sphere}
\label{subsec:Incompete}

Consider the inviscid, incompressible, steady flow around a solid sphere.
We assume that the incident flow is homogeneous, unidirectional, and subsonic, so the entire flow is irrotational.
Here, the mass density $\rho=\init{\rho}$ is constant, the radius of the sphere is taken (henceforth) as $R=1$, and the constant $\initial{\bm{v}}=-\initial{u}\unit{z}$ is taken on $\myH$ far upstream, $\init{z}\to\infty$.
It is thus natural to choose $\mygconst{\rho}=\init{\rho}$, $\mygconst{R}=1$ and $\mygconst{v}=\init{u}$, so $\myshort=\init{\rho}\init{u}$.

As $\grad\cdot\bm{v}=0$ and $\grad\times\bm{v}=0$, the flow satisfies the Poisson equation $\grad^2 \Phi=0$ for the velocity potential $\Phi$, defined by $\bm{v}=\init{u}\grad \Phi$.
The solution for boundary conditions of a vanishing normal velocity across the sphere is
\begin{equation}\label{eq:incopote_Phi}
\Phi=-z\left(1+\frac{1}{2r^3}\right) \fin
\end{equation}
Here we primarily use cylindrical coordinates, but occasionally allude to spherical coordinates $\{r,\theta,\phi\}$, where
$r^2=\myrho^2+z^2$. Thus,
\begin{align} \label{eq:PotIncFlow}
& v_z = \init{u}  \left(-1-\frac{1}{2r^3} +\frac{3z^2}{2r^5}\right) \quad \mbox{and} \quad v_\myrho = \frac{3\init{u}\myrho z}{2r^5} \fin
\end{align}
The stream functions can be chosen as $\lambda_1=\phi$ and, according to Equation~(\ref{eq:lambdas_v}),
\begin{equation}\label{eq:lambda_incopote}
\lambda \equiv \lambda_2 = \frac{\myrho^2}{2}\left(1-\frac{1}{r^3}\right)\coma
\end{equation}
such that $\init{\myrho}=(2\lambda)^{1/2}$.

Solving Euler's Equation~(\ref{eq:Euler}) gives the pressure,
\begin{equation}
P=\init{P}+\init{\rho}\init{u}^2\frac{4r^3-5+3\left(4r^3-1\right)\cos2\theta}{16 r^6}\coma
\end{equation}
where $\cos2\theta=(z^2-\myrho^2)/r^2$.
Note that an incident pressure smaller than $(5/8)\init{\rho}\init{u}^2$ would lead to a nonphysical $P<0$ around $z=0$, as the plasma would be too cold to remain incompressible.

Consider the evolution of a magnetic field, initially given by $\init{\bm{B}}=(\init{B}_\myrho,\init{B}_\phi,\init{B}_z)$, for different choices of $q$. In this subsection, round-bracket vectors are in cylindrical coordinates.

\begin{figure*}[t!]
\centering
\includegraphics[width=1\textwidth]{\myfig{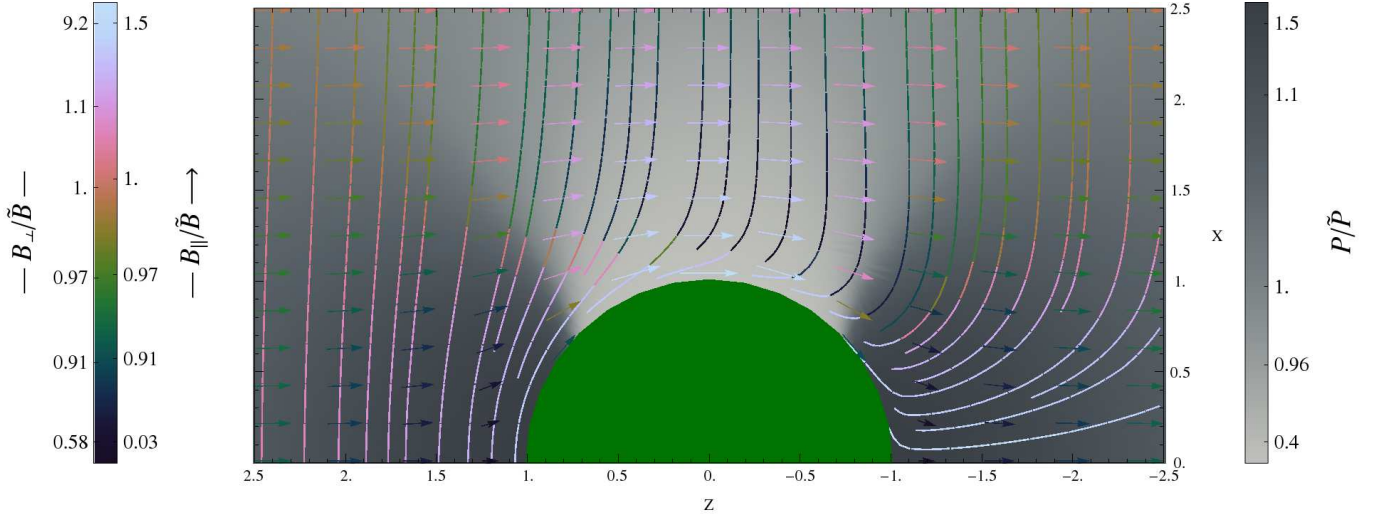}}
\caption{Potential, incompressible, steady flow (arrows with direction and length proportional to $\bm{v}$) around a solid sphere (green half disk), for a uniform magnetic field far upstream.
The constant $\phi$ slice shows the pressure enhancement (gray scale; for $\init{P}=\init{\rho}\init{u}^2$) and the magnetic amplification $B/\init{B}$ \citep[colorscale;][]{Green2011} for field lines parallel (arrows) or initially perpendicular (solid curves) to the flow.}
\label{fig:PotIncSphere1}
\end{figure*}

\subsubsection{The natural choice $\mythird=z$}

For the choice $\mythird=z$, which is natural because $z$ monotonically decreases for all fluid elements,
the magnetic field is given by (\cf Equation \ref{eq:B_Cylindrical})
   \begin{align}\label{eq:B_incopote}
\bm{B}=\init{B}_\phi\frac{\myrho}{\init{\myrho}}\unit{\phi}-\init{B}_\myrho\frac{\init{\myrho}\init{u}}{\myrho v_z}\unit{\myrho}-\frac{\bm{v}}{\init{u}}\left(\init{B}_z+\init{B}_\myrho\init{\myrho}\init{u}\frac{\partial\Delta t}{\partial\lambda}\right)\coma
     \end{align}
where
   \begin{align}\label{eq:Fz_of_incopote}
     \frac{\partial \deltats}{\partial\lambda}& = \left(\frac{\partial \deltats}{\partial\lambda}\right)_{\phi,z,\init{z}}
      =-\frac{12}{\init{u}}\int\limits_{\infty}^{z}\frac{r'^8\left(r'^2-5z'^2\right)dz'}{\left(r'^2+2r'^5-3z'^2\right)^3}\fin
     \end{align}
The integral is taken at a constant $\lambda$, so $r'(\lambda,z')$ is found from Equation~(\ref{eq:lambda_incopote}).

Figure~\ref{fig:PotIncSphere1} shows a constant $\phi$ slice through the flow, depicting the pressure (grayscale) and the magnetic field components parallel (colorscaled arrows) or initially perpendicular (colorscaled curves) to the velocity.
For illustrative purposes, in the figure we take $\init{P}=\init{\rho}\init{u}^2$.

The electric field here is a special case of Equation~(\ref{eq:E_Cylindrical}), namely
\begin{align}\label{eq:E_incopote}
\bm{E}
  =\frac{\myrho}{ \init{\myrho}c}\begin{pmatrix}[1.7]
        v_z\init{B}_\phi \\
       \frac{2\lambda}{\myrho^2}\,  \init{u}\init{B}_\myrho \\
       - v_\myrho\init{B}_\phi
  \end{pmatrix}
    \fin
\end{align}

\subsubsection{The problematic choice $\mythird=r$}

Another seemingly possible choice is $\mythird=r$.
According to Equation~(\ref{eq:vecB_Deltaq}), the magnetic field may then be written as
      \begin{align}\label{eq:B_incopote_vec}
     \bm{B}=\init{B}_\phi\frac{\myrho}{\init{\myrho}}\unit{\phi}+\init{B}_\myrho\frac{\myrho r  }{\init{\myrho}z}\unit{\theta}-\frac{\bm{v}}{\init{u}}\left(\init{B}_z+\init{B}_\myrho\init{\myrho}\init{u}\frac{\partial\Delta t}{\partial\lambda}\right),
          \end{align}
where
\begin{flalign}\label{eq:F_of_incopote}
\frac{\partial \deltats}{\partial\lambda} = \left(\frac{\partial \deltats}{\partial\lambda}\right)_{\phi,r,\init{r}}
  =\int\limits_{\infty}^{r}\frac{ -r'^4dr'/\init{u}}{\left({r'^3-1}\right)^2\left(1-\frac{2\lambda r'}{r'^3-1}\right)^{\frac{3}{2}}}\fin
\end{flalign}

However, this result, obtained previously \citep{DursiPfrommer08,Romanelli14} by solving the MHD equations, is strictly valid only in the half space $\theta<\pi/2$.
At the surface $\theta=\pi/2$, the denominator of the integrand diverges, and the integration cannot be continued.
This is to be expected, as surfaces of constant $\lambda$ and surfaces of constant $r$ overlap at $\theta=\pi/2$, as $\grad\lambda$ and $\grad r$ become parallel.
The problem can be circumvented by computing the converging $\Delta t$,
rather than its diverging derivative.

The integral in Equation~(\ref{eq:F_of_incopote}) can be solved analytically as a power series in $\lambda$; see appendix \ref{sec:AppPerpBPotIncFlow}.

\subsubsection{The orthogonal choice $\mythird=\Phi$}

Another optional choice for $\mythird$ is the velocity potential $\Phi$. Here, the reciprocal basis becomes
\begin{align}
\mya_1=\myrho\,\hat{\bm{\phi}}\pcoma\quad \mya_2=\frac{\init{u}}{\myrho v}\left(\hat{\bm{v}}\times\unit{\phi}\right)\pcoma\quad \mya_3=\hat{\bm{v}}\fin
\end{align}
These vectors are perpendicular to one another, and therefore may be useful as an orthogonal (albeit not orthonormal) basis. The magnetic field here is given by
\begin{align}\label{eq:B_Phi}
\bm{B}  =
\begin{bmatrix}[1.7]
       \init{\myrho}^{-1}\init{B}_\phi \\
       \init{\myrho} \init{B}_\myrho\\
      -\frac{v}{\init{u}}\init{B}_z
  \end{bmatrix}- \begin{bmatrix}[1.7]
        0 \\
        0\\
       \init{\myrho} v
    \end{bmatrix}\init{B}_\myrho\frac{\partial\Delta t}{\partial\lambda} \coma
\end{align}
where
\begin{align}
  &\frac{\partial \deltats}{\partial\lambda} = \left(\frac{\partial \deltats}{\partial\lambda}\right)_{\phi,\Phi,\init{\Phi}}\\\nonumber
  &=\frac{48}{\init{u}}\int\limits_{-\infty}^{\Phi}\frac{ z'^4+4r'^8\Phi'^2+10z'^4r'^3\left(1-2r'^3\right)}{r^{-11}z^{-4}\left[4r'^8\Phi'^2+3z'^4\left(1-4r'^3\right)\right]^3}
\,d\Phi'\fin
  \end{align}
The parameters $r'(\lambda,\Phi')$ and $z'(\lambda,\Phi')$ are found from
Equations~(\ref{eq:incopote_Phi}) and (\ref{eq:lambda_incopote}).

\subsection{Stokes Incompressible Steady Flow Around a Sphere}
\label{subsec:Stokes}

\begin{figure*}[t!]
\centering
\includegraphics[width=1\textwidth]{\myfig{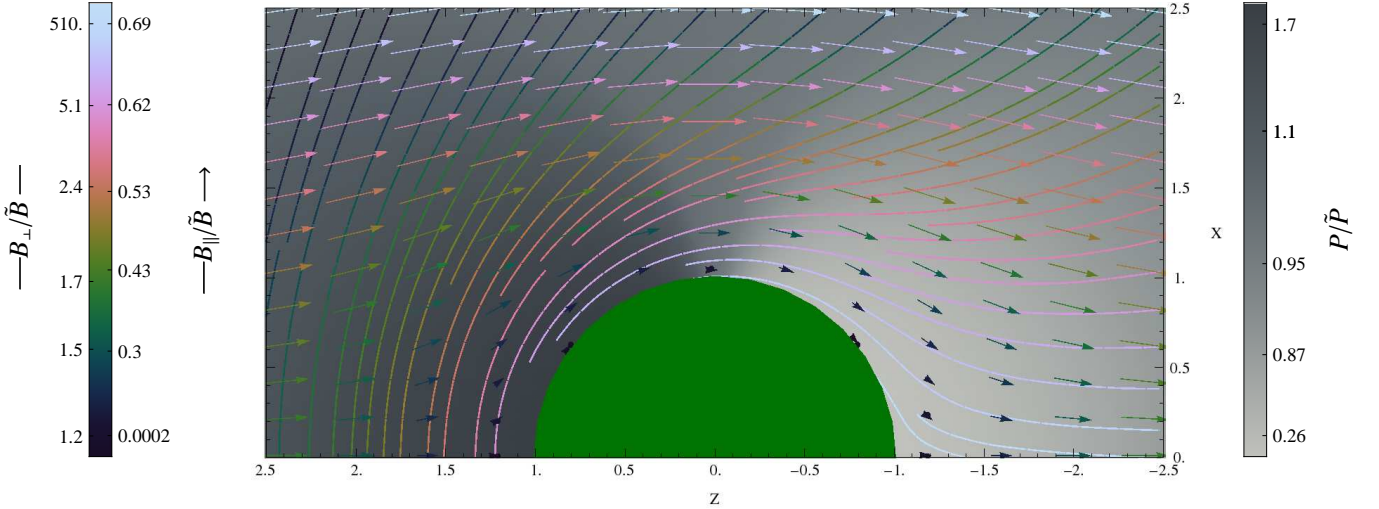}}
\caption{Stokes, incompressible, steady flow around a solid sphere. Notations are as in Figure~\ref{fig:PotIncSphere1},  with $\init{P}=2\init{\rho}\init{u}^2/\Rey$.}
\label{fig:stokesSphere1}
\end{figure*}

Consider the Stokes, incompressible, steady flow around a solid sphere.
As in Section~\ref{subsec:Incompete}, $\rho=\init{\rho}$ is constant, the radius of the sphere is taken as $R=1$, we assume that the incident flow $\initial{\bm{v}}\equiv-\initial{u}\unit{z}$, taken on $\myH$ far upstream ($\init{z}\to\infty$), is homogeneous and unidirectional, and use a mixture of cylindrical and spherical coordinates.
It is again natural to choose $\mygconst{\rho}=\init{\rho}$, $\mygconst{R}=1$ and $\mygconst{v}=\init{u}$, so $\myshort=\init{\rho}\init{u}$.

Stokes flows are characterized by viscous forces much greater than the inertial forces, so
 the flow is governed by the approximate Navier-Stokes equation
 \begin{equation}\label{eq:stokes_NSE}
 0=-\grad\left(\frac{P}{\initial{\rho}}\right)+\nu\grad^2\boldsymbol{v},
 \end{equation}
where $\nu$ is the kinematic viscosity, assumed constant.
Taking the curl of this equation gives
 \begin{equation}
 \grad\times\left(\grad^2\bm{v}\right)=0\fin
 \end{equation}
The solution, for boundary conditions of a hard sphere, is
 \begin{align}
 v_z&=-\init{u}\left(1-\frac{1}{r}\right)\left[1+\frac{\left(r^2-3z^2\right)\left(r+1\right)}{4r^4}\right]\coma
 \end{align}
  and
   \begin{align}
  v_\myrho&=3\init{u}\left(1-\frac{1}{r^2}\right)\frac{\myrho z}{4r^3}\fin
 \end{align}

Solving Equation~(\ref{eq:stokes_NSE}) gives the pressure distribution,
  \begin{equation}
P=\init{P}+\init{\rho}\init{u}\nu\frac{3z}{2r^3}\fin
  \end{equation}
Here too, the incident plasma cannot be too cold, as an incident pressure $\init{P}<(3/2)\init{\rho}\init{u}^2/\Rey$, where $\Rey=\init{u}R/\nu$ is the incident Reynolds number, would lead to a nonphysical $P<0$ around the $z<0$ axis.

The stream functions can be chosen as $\lambda_1=\phi$ and, according to Equation~(\ref{eq:lambdas_v}),
\begin{equation}\label{eq:lambda_stokes}
\lambda \equiv \lambda_2=\frac{\myrho^2}{2}\left(1-\frac{1}{r}\right)^2\left(1+\frac{1}{2r}\right)\fin
\end{equation}
As in the case of a potential flow, here too $\init{\myrho}=(2\lambda)^{1/2}$, and we proceed similarly with the natural choice $\mythird=z$.
The magnetic field is given by
   \begin{align}\label{eq:B_Stokes}
\bm{B}=\init{B}_\phi\frac{\myrho}{\init{\myrho}}\unit{\phi}-\init{B}_\myrho\frac{\init{\myrho}\init{u}}{\myrho v_z}\unit{\myrho}-\frac{\bm{v}}{\init{u}}\left(\init{B}_z+\init{B}_\myrho\init{\myrho}\init{u}\frac{\partial\Delta t}{\partial\lambda}\right) \coma
     \end{align}
which is formally equivalent to Equation~(\ref{eq:B_incopote}). Here,
 \begin{align}\label{eq:Fz_of_stokes}
\frac{\partial \deltats}{\partial\lambda}& = \left(\frac{\partial \deltats}{\partial\lambda}\right)_{\phi,z,\init{z}}
 \\\nonumber
 &=\frac{48}{\init{u}}\int\limits_{\infty}^{z}
 \frac{r'^6-5z'^2r'^2+r'^4\left(1+3z'^2\right)}
 {\left(1-\frac{1}{r'^2}\right)^3\left(r'^2-3z'^2+\frac{4r'^4}{1+r'}\right)^3}\,dz'\coma
 \end{align}
where $r'(\lambda,z')$ is found from Equation~(\ref{eq:lambda_stokes}).

Figure~\ref{fig:stokesSphere1} shows the pressure and magnetization in such a flow, with the same notations used in Figure~\ref{fig:PotIncSphere1}.
For illustrative purposes, in the figure $\init{P}=2\init{\rho}\init{u}^2/\Rey$.

Here too, the electric field is a special case of Equation~(\ref{eq:E_Cylindrical}), and is given by Equation~(\ref{eq:E_incopote}).

\subsection{One-dimensional, Time Dependent Flow}
\label{sec:rarefaction_wave_Eample}

As a simple illustration of magnetization in the time-dependent regime, consider the case of an arbitrary 1D flow.
Classical examples include the plasma in a cylindrical pipe, disturbed by a moving piston, or a 1D rarefaction wave.
If discontinuities are present in the flow, the following applies separately to each region bounded by surfaces of discontinuity or $\myH$.

Without loss of generality, take the flow in the $\hat{\bm{x}}$ direction, such that $\bm{v}=v\hat{\bm{x}}$ and the perpendicular velocity components vanish.
A simple choice of path functions is then
\begin{align}
\Lambda_2&=y\quad \mbox{and} \quad \Lambda_3=z\coma
\end{align}
as both are constant along the flow.
Equations (\ref{eq:pathv}) and (\ref{eq:pathrho}) therefore constrain $\Lambda_1$, as
\begin{align}\label{eq:1DdLdx}
\left(\frac{\partial \Lambda_1}{\partial x}\right)_{y,z,t}=-\frac{\rho}{\mygconst{\rho}}
\quad \mbox{and} \quad
 \left(\frac{\partial \Lambda_1}{\partial t}\right)_{x,y,z}=\frac{\rho v}{\mygconst{\rho}}\fin
\end{align}
The solution is the path function family
\begin{equation}\label{eq:Lambda1_1D}
\Lambda_1(x,t)=-\frac{1}{\mygconst{\rho}}\int^x\rho(x',t)\, dx' + \mathcal{C} \coma
\end{equation}
where we used the continuity equation, and $\mathcal{C}$ is an insignificant constant.

For a given $\init{\bm{l}}=(\init{l}_x,\init{l}_y,\init{l}_z)$, Equation~(\ref{eq:initlpath}) yields the Lagrangian constants
\begin{align}
\ilLambda_1=-\frac{\init{\rho}}{\mygconst{\rho}}\init{l}_x \pcoma \quad
\ilLambda_2=\init{l}_y \pcoma \quad
\ilLambda_3=\init{l}_z \coma
\end{align}
where here and in the rest of Section~\ref{sec:rarefaction_wave_Eample}, round brackets are in Cartesian coordinates.
Equivalently, for $\init{\bm{B}}=(\init{B}_x,\init{B}_y,\init{B}_z)$, Equation~(\ref{eq:initBpath}) gives
\begin{align}
\init{B}_1=-\frac{\init{\rho}}{\mygconst{\rho}}\init{B}_x \pcoma \quad
\init{B}_2=\init{B}_y \pcoma \quad
\init{B}_3=\init{B}_z \fin
\end{align}

Finally, Equations~(\ref{eq:l_patht}) and (\ref{eq:B_patht}) give, respectively
\begin{equation}\label{eq:l_rarefaction}
\bm{l}=\left( \frac{\init{\rho}}{\rho}\init{l}_x, \init{l}_y, \init{l}_z\right)\coma
\end{equation}
and
\begin{equation}\label{eq:B_rarefaction}
\bm{B}=\left( \init{B}_x, \frac{\rho}{\init{\rho}}\init{B}_y, \frac{\rho}{\init{\rho}}\init{B}_z \right) \fin
\end{equation}

Equations (\ref{eq:l_rarefaction}) and (\ref{eq:B_rarefaction}) can be seen directly from Lagrangian considerations.
Since the gas flows only in the $x$ direction, the perpendicular (to the flow) distance between fluid elements is constant.
Perpendicular length elements $\{l_y,l_z\}$ and the parallel magnetic field $B_x$, are thus conserved.
The parallel distance between nearby fluid elements is simply $l_x\propto \rho^{-1}$, so the perpendicular fields follow $\{B_y,B_z\}\propto\rho$.

\section{Astrophysical examples}
\label{sec:Astrophysical applications}

The simplicity of the above MHD formalism allows us to easily obtain results, both general and specific, for the evolution of electromagnetic fields in a wide range of astronomical systems.
In particular, the formalism yields analytic expressions for the electric field, directly incorporates time-dependent flows, and admits inhomogeneous initial conditions in both space and time, $\init{\bm{B}}=\init{\bm{B}}(\bm{x};t)$, whether or not the flow itself is time-dependent.

In Section~\ref{sec:spacetime_Btilde}, we demonstrate such inhomogeneous scenarios by considering substructure in $\init{\bm{B}}$.
In particular, some observational consequences are outlined for a stellar wind behind a planet or a moon.
Near the nose of a body, the magnetic field is greatly amplified; the resulting magnetic layer, analyzed in Section~\ref{sec:Boundarylayer}, becomes dynamically significant and distorts the flow.
In Section~\ref{sec:ionospheric electric field}, we derive analytic expressions for the electric field above heliospheres and magnetospheres.
\fixapj{
Sophisticated astronomical scenarios typically require a numerical computation; the integration of our method into numerical simulations is discussed in Section~\ref{sec:Discussion}.
}

\begin{figure*}[t!]
\centering
\includegraphics[width=1\textwidth]{\myfig{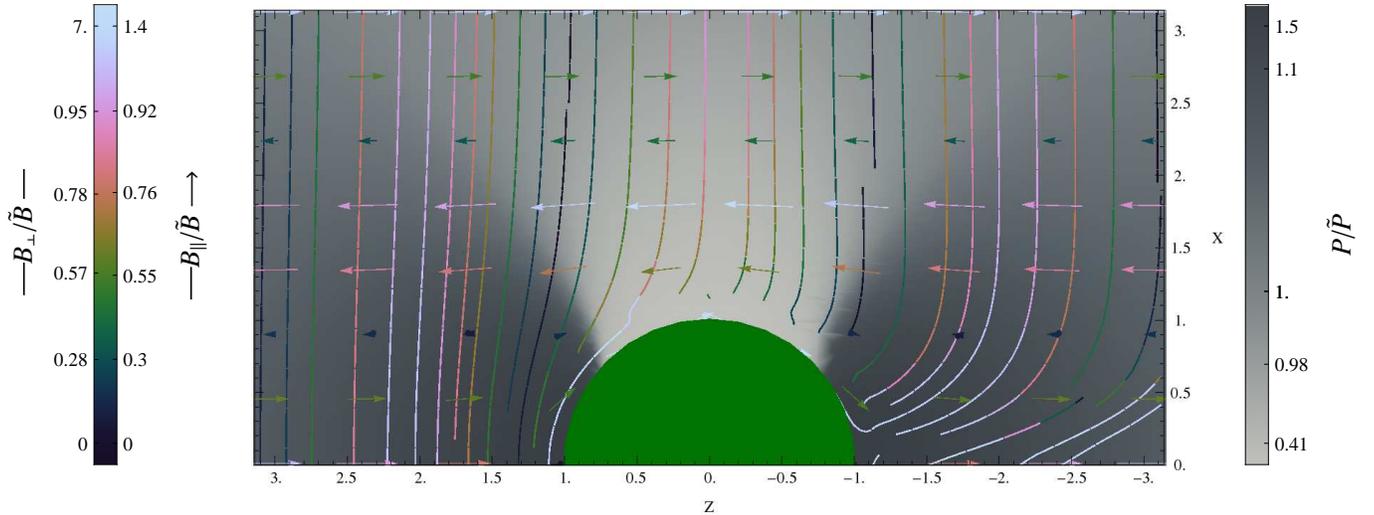}}
\caption{Evolution of a magnetic Fourier component, in an incompressible potential steady flow around a solid sphere. Notations are as in Figure~\ref{fig:PotIncSphere1}; here arrows are either along or opposite to the flow. The magnetic field far upstream of the sphere is given by Equation~(\ref{eq:timespaceinitB}) with $k=2$, $\omega=\pi$, and $t=0$.
}
\label{fig:PotIncSphereTSB}
\end{figure*}

\subsection{Spacetime Dependent $\init{\bm{B}}$}
\label{sec:spacetime_Btilde}

Given a flow and an initial magnetic configuration $\init{\bm{B}}_{in}$ on $\myH$, the electromagnetic field solution in Equations~(\ref{eq:vecB_Deltaq}), (\ref{eq:ESteady}), (\ref{eq:vecB_DeltaaT}), (\ref{eq:B_pathfour}), (\ref{eq:B_patht}) or (\ref{eq:Etimedependent}), depends only on knowing the Lagrangian constant $\init{\bm{B}}$ corresponding to every spacetime location.
For an initially uniform magnetic field, $\init{\bm{B}}$ is everywhere constant, and the result is straightforward even for time-dependent flows.
For a spacetime-dependent initial magnetic field, this mapping can become complicated, especially for time-dependent flows.

Consider however the case of a steady flow, with the choice $q=\Delta t$.
Here, even if $\init{\bm{B}}_{in}(\bm{x},t)$ varies in space and time, we may directly determine $\init{\bm{B}}$ using
\begin{align}\label{eq:Binit_spacetime}
\init{\bm{B}}(\bm{x};t) & = \init{\bm{B}}(\lambda_1,\lambda_2,\Delta t;t)=\init{\bm{B}}(\lambda_1,\lambda_2,0;\init{t}\equiv t-\Delta  t) \nonumber \\
&  = \init{\bm{B}}_{in}(\lambda_1,\lambda_2,0;\init{t})
\coma
\end{align}
where $\init{t}$ is the time at which the fluid element, which presently lies at $(\bm{x},t)$, was on $\myH$.
Equation (\ref{eq:Binit_spacetime}) applies for any choice of $q$, if Equation~(\ref{eq:time_t}) is used to compute the travel time.
For example, in cylindrical coordinates with $\myH$ at $z\to \infty$, we may compute it by
 \begin{equation}
\Delta t=\int\limits_{\infty}^{z}\frac{dz'}{v_z} \label{eq:ztravel time}\fin
 \end{equation}

An arbitrary unperturbed $\init{\bm{B}}(\bm{x},t)$ can be Fourier decomposed into components of wavenumber $k$ and angular frequency $\omega$.
As a simple example, assume a flow in the $(-z)$ direction, and consider a Fourier component of the form
\begin{equation}\label{eq:timespaceinitB}
\init{\bm{B}}=B_0(\cos(\omega\init{t}),\,0,\,\cos(k\init{\myrho}))\coma
\end{equation}
where $B_0$ is a constant, and round-bracket vectors in this subsection are in Cartesian coordinates.
Notice that no spatial component of this $\init{\bm{B}}$ contributes to Gauss' law, and that a $t$-dependence on $\myH$ induces a $z$-dependence in its vicinity, \ie Equation~(\ref{eq:timespaceinitB}) becomes
\begin{equation}\label{eq:initBH}
\init{\bm{B}}=B_0(\cos[\omega(t+z/\init{v})],\,0,\,\cos(k \myrho))
\end{equation}
in the region where the flow is approximately uniform.

Figure.~\ref{fig:PotIncSphereTSB} presents the evolution of the Fourier component Equation~(\ref{eq:timespaceinitB}), in an incompressible potential flow around a sphere (see Section~\ref{subsec:Incompete}).
As Figures~\ref{fig:PotIncSphereTSB} and \ref{fig:IPIPRL} show, substructure in the far upstream magnetic field is non-trivially imprinted on the downstream flow. In situ measurements of $\bm{B}$ downward of an object such as a planet would thus be sensitive to both the substructure and details of the flow.

As an example, consider measurements of the inverse polarity reversal layer (henceforth IPRL) induced by a planetary body \citep[][]{Romanelli14}. The IPRL is defined as the layer in which the magnetic field component in the direction of the initial flow changes its sign, implying in our case that $B_z=0$.
According to Equation~(\ref{eq:B_Cylindrical}), for $\mythird=z$, the IPRL is described for a general flow by the simple formula
\begin{equation}
\init{B}_z = \init{B}_2 \init{v}\frac{\partial \Delta t}{\partial \lambda} = \frac{\init{v}\left({\init{\mymu}_\myrho\init{B}_z-\init{\mymu}_z\init{B}_\myrho}\right)}{{\myshort}}\frac{\partial\deltats}{\partial\lambda}
\label{eq:IPRL}\fin
\end{equation}

Consider the IPRL for a magnetic field with initial oscillatory substructure.
For an initial magnetic inclination $\theta_0$, let
\begin{equation}\label{eq:initBIPRL}
\init{\bm{B}}=B_0\left(\sin\theta_0,\,0,\,\cos\theta_0\right)+\hat{\bm{x}}\myepsilon B_0 \sin\left(\theta_0\right)\cos\omega\init{t}\coma
\end{equation}
where $\myepsilon<1$ is a small number; this $\init{\bm{B}}$ trivially satisfies Gauss' law near $\myH$.
A snapshot of the resulting magnetic configuration is shown in Figure~\ref{fig:IPIPRL}, where for illustrative purposes we adopt the incompressible, steady potential flow of Section~\ref{subsec:Incompete}.
As the figure shows, the far upstream magnetic substructure with wavelength $2\pi \tilde{u}/\omega$ is directly imprinted on the IPRL.

Such a model may describe the magnetization not only around planets and moons, but also around stellar systems moving through the ISM, around clumps and bubbles moving through the IGM, etc.

\subsection{Magnetic Layers}
\label{sec:Boundarylayer}

\begin{figure*}[t!]
\centering
\includegraphics[width=1\textwidth]{\myfig{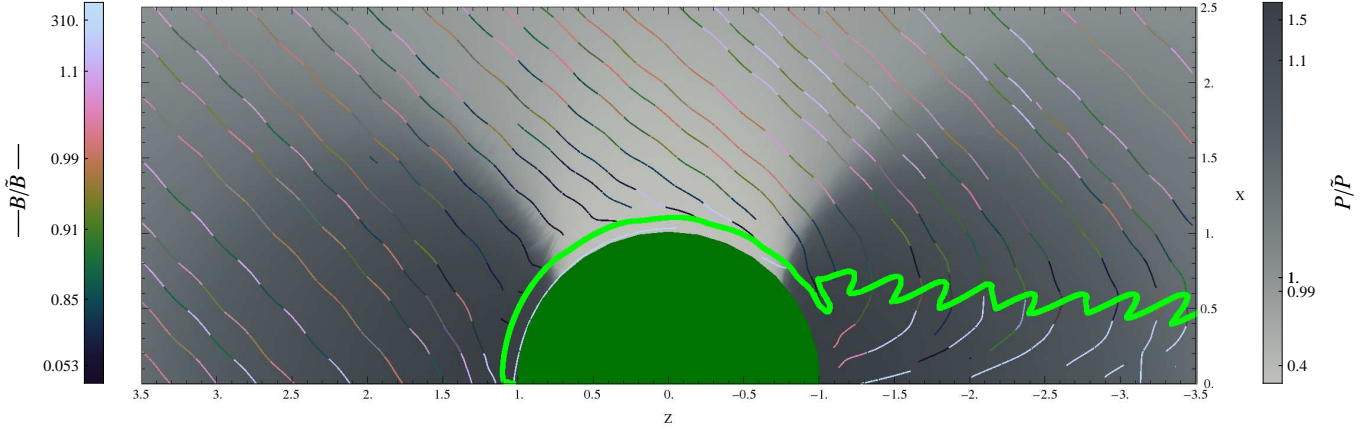}}
\caption{Inverse polarity reversal layer (IPRL; defined by $B_z=0$; thick green curve) for substructure in the magnetic field far upstream of the sphere. Notations and flow are as in Figure~\ref{fig:PotIncSphere1}. Here the magnetic field far upstream of the sphere is given by Equation~(\ref{eq:initBIPRL}), where $\theta_0=\pi/4$, $\myepsilon =0.25$, $\omega=6\pi$, and $t=0$.
}
\label{fig:IPIPRL}
\end{figure*}

Under our assumptions, as the flow approaches the stagnation point in front of an object, the magnetic field initially perpendicular to the flow diverges, as seen in Figures~\ref{fig:PotIncSphere1}--\ref{fig:IPIPRL}.
The divergence here, around the object, and in the wake behind it, may be regarded as due to the accumulation of infinitely many field lines, continuously stretched around the object, eventually wrapped closely around it and stretching to infinity along the wake.

These nonphysical divergencies signal the breakdown of our assumptions in the above regions. In particular, the magnetic field here can no longer be regarded as dynamically insignificant, and eventually resistivity can no longer be neglected \citep[\eg][]{ChackoHassam97, Lyutikov06}.

It is useful to consider the extent of the highly magnetized layer forming in front of an object in an arbitrary axisymmetric flow.
Such layers form as the magnetized solar wind drapes around a planetary magnetopause, as the magnetized ISM drapes around a bubble or a stellar heliopause, and as the magnetized IGM drapes around a large scale-clump or bubble.
The regions surrounding such objects could involve plasma depletion layers with dynamically significant magnetic fields, and lead to plasma instabilities, magnetic reconnection, and turbulence \citep[\eg][]{Dibraccio_etal2011,Sulaiman_etal2014}.
In the following, we neglect the small modifications to the flow far from the object due to the presence of a highly magnetized region.	 

 Define the magnetic pressure fraction
\begin{equation}
\epsilon_B\equiv\beta_p^{-1}=\frac{B^2/8\pi}{\myPT} \coma
\end{equation}
where $\beta_p$ is the beta of the plasma, and $\myPT$ is the thermal pressure.
The fractional magnetic amplification may then be defined as
\begin{equation}\label{eq:Bamplification}
\frac{\epsilon_B}{\initial{\epsilon}_{B}} = \left(\frac{B}{\initial{B}}\right)^2\left(\frac{\myPT}{\myPTi}\right)^{-1}\fin
\end{equation}

\begin{figure*}[t!]
\centering
\includegraphics[width=0.4\textwidth]{\myfig{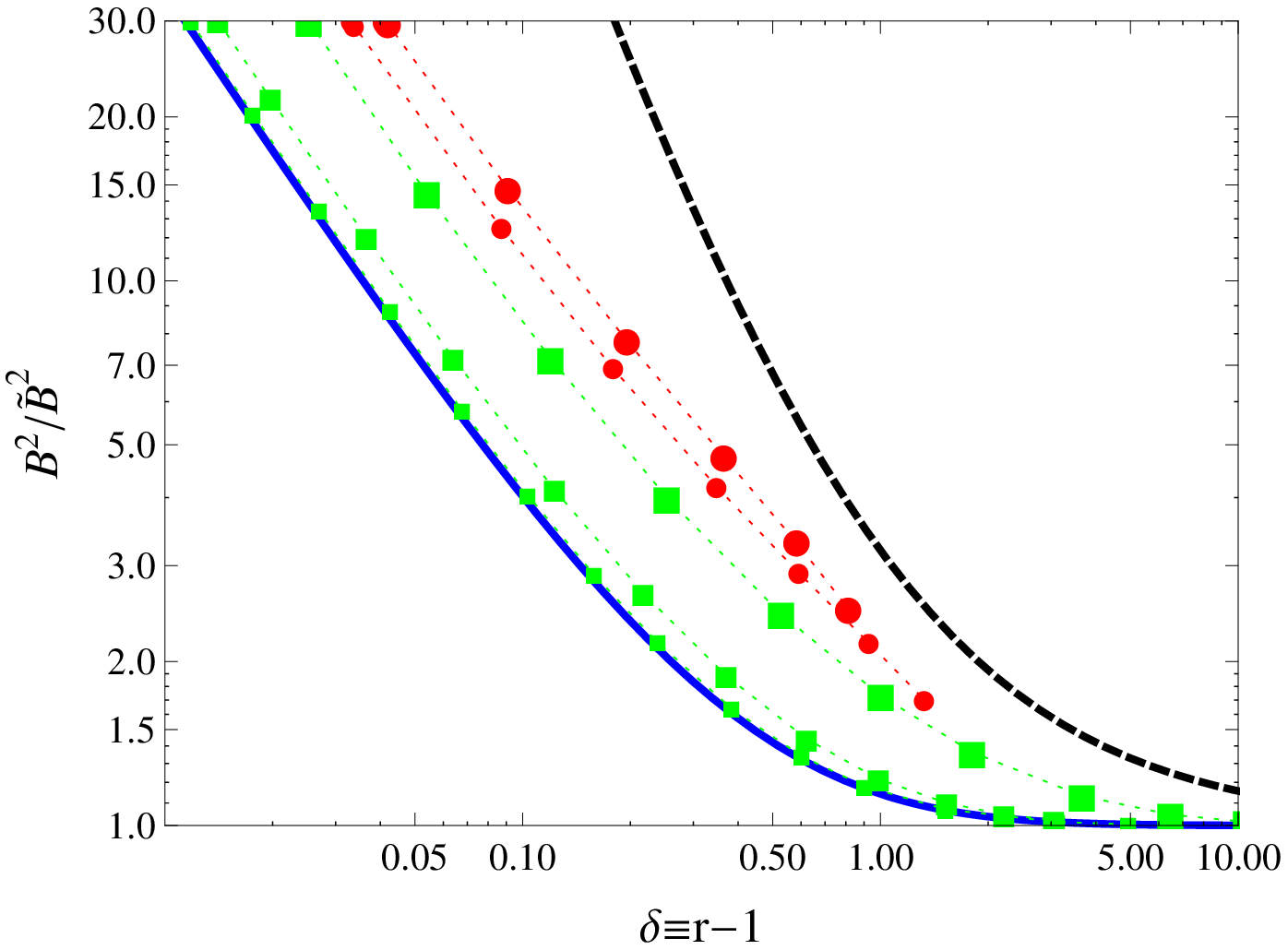}}
\includegraphics[width=0.4\textwidth]{\myfig{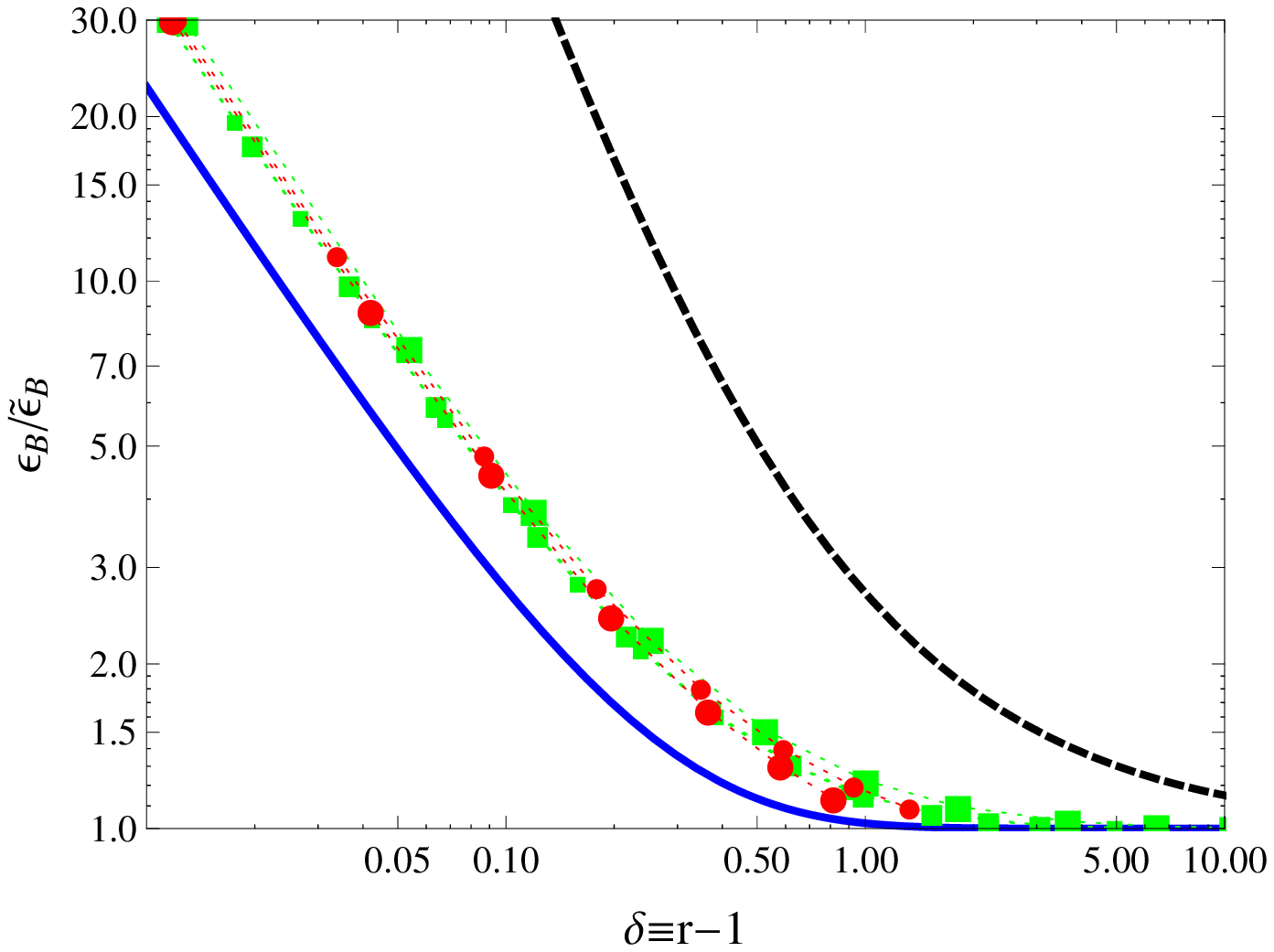}}
\caption{Head on ($\theta=0$) amplification of the magnetic pressure (left) and magnetic pressure fraction (right), plotted against the distance $\delta$ from the nose of the unit sphere, for \fixapj{fields initially perpendicular to} various steady flows: incompressible potential (solid blue; for $\init{P}=\init{\rho}\init{v}^2$), Stokes incompressible (dashed black; for $\init{P}=2\init{\rho}\init{v}^2/\Rey$), and inviscid compressible (symbols with dotted lines to guide the eye) in the subsonic ($\initial{M}\in\{0.2, 0.5, 0.95\}$; green squares, small to large) and mildly supersonic ($\initial{M}\in\{1.2,1.4\}$; red circles, small to large) regimes.}
\label{fig:amplificaion_distance_all1}
\end{figure*}

Figure~\ref{fig:amplificaion_distance_all1} shows the amplifications of magnetic pressure $(B^2/\init{B}^2)$ and pressure fraction $(\epsilon_B/\init{\epsilon}_B)$, as a function of the distance $\delta\equiv r-1$ in front of the unit sphere, for various steady flows: an incompressible potential flow, a Stokes incompressible flow, and inviscid compressible flows in the subsonic and mildly supersonic regimes.
As the above flows are approximately (for the mildly supersonic case), or precisely, irrotational along the axis, Equations~(\ref{eq:AxialB}) and (\ref{eq:AxialC}) become
\begin{align}\label{eq:B_onaxis}
B(z)^2&=\frac{\rho/u}{\init{\rho}/\init{u}} \initial{B}_{\perp}^2+
\frac{\myj^2}{\initial{\myj}^2}\initial{B}_{\parallel}^2 \coma
\end{align}
where $\initial{B}_{\perp}$ and $\initial{B}_{\parallel}$ are the axial magnetic field components initially perpendicular and parallel to the flow.
\fixapj{The figure focuses on the diverging, perpendicular component.}

As Figure \ref{fig:amplificaion_distance_all1} shows, viscosity induces magnetic amplification farther from the body.
Compressibility has a similar, but somewhat weaker, effect on magnetization, but the magnetic pressure fraction is not strongly affected. The large difference in $\epsilon_B$ between the incompressible and compressible inviscid flows arises from the different pressure profiles corresponding to the different equations of state.

Close to stagnation, the flow is adiabatic, $\rho$ and $\myPT$ approach constant values, and $\bm{B}$ becomes primarily transverse, so
\begin{equation} \label{eq:StagnationB}
\epsilon_B\propto \frac{B_{\perp}^2}{\myPT} \propto B_\perp^2 \propto \frac{\rho^2}{j} \propto u^{-1} \coma
\end{equation}
which scales as $\delta^{-2}$ for the Stokes flow, and as $\delta^{-1}$ for the incompressible potential flow; the latter  approximately characterizes the stagnation region even for compressible flows, except in the highly supersonic regime.

Magnetic amplification beyond $\epsilon_B\simeq 1$, say, is not self-consistent, as our passive magnetization assumption fails.
Instead, one expects a highly magnetized, self-regulating layer to form; its width $\delta$ depends on the initial magnetization level $\init{\epsilon}_B$.
For simplicity, consider an initially hot plasma, such that $\myPTi\gg\init{\rho}\init{u}^2$ for an inviscid flow, or $\myPTi\gg \init{\rho}\init{u}^2/\Rey$ for the Stokes flow.
In this regime $\myPT\simeq\myPTi$, so according to Equation~(\ref{eq:StagnationB})
\begin{equation} \label{eq:ApproxBWidth}
\delta \sim \begin{cases}
\init{\epsilon}_B & \mbox{for negligible viscosity;} \\
\init{\epsilon}_B^{1/2} & \mbox{for strong viscosity.}
\end{cases}
\end{equation}

For the compressible flows shown in Figure~\ref{fig:amplificaion_distance_all1}, we use the approximate solutions of \citet[][using $\gamma=5/3$ and $\beta=0.52$]{KeshetNaor2014}.
In the supersonic regime, $\myH$ is assumed to be upstream, so the jumps in $B_\perp$ and $\myPT$ across the shock are taken into account.
Here, the flow remains nonmagnetized until reaching the shock, at the so called standoff distance $\delta=\Delta$; for high $\init{M}_0$, the magnetized layer slightly shrinks as $\Delta$ diminishes.

\subsection{$\bm{E}$ Above Heliospheres and Magnetospheres}
\label{sec:ionospheric electric field}

The electric field in ideal MHD was shown to be a simple function of the stream functions (see Equations~(\ref{eq:ESteady)}, (\ref{eq:Etimedependent}), (\ref{eq:E_Cylindrical})), so it is easy to infer $\bm{E}$ directly in various astronomical circumstances.
As an example, in what follows we consider the electric field outside a heliosphere (of some star) or magnetosphere (of some stellar object).
We approximate the bow shock as weak \citep[or absent; for the possibility that there is no bow shock in front of the solar system, see][]{McComasEtAl2012}, and model the discontinuity (heliopause or magnetopause) using a Rankine half-body.
These assumptions may be marginally invalid for the magnetopause of Earth, but roughly apply for the Sun's heliosphere, and may be valid for other objects such as the induced magnetospheres of Venus and Mars.

The Rankine half-body model was applied to the Sun's heliopause by \citet{Parker61}.
Models for Earth's magnetopause are more sophisticated \citep[\eg][ and references threin]{Fairfield1971,Tsyganenko1989,Shue_etal1997,ChaoEtAl2002,LinEtAl2010,LuEtAl2011,Tsyganenko2013}, and better calibrated by satellite observations.
However, no present analytic model captures the full complexity of the magnetopause structure, including the time dependence of the geodipole tilt direction and the stellar wind variability \citep[][]{Olson1969,ZhouRussell1997}, so for simplicity we resort to Equation~(\ref{eq:RankineHalfSphere}).

The normalized potential of the Rankine half-body is
\begin{equation} \label{eq:RankineHalfSphere}
\Phi = -\left( z+\frac{\myPLS}{r} \right) \coma
\end{equation}
where mixed coordinates are used, as in Section~\ref{subsec:Incompete}, and $L_m$ is the distance between the object (star, planet, or moon) and the nose of the discontinuity.
This simplified potential flow is axisymmetric and incompressible.
It approximates the flow ahead of the discontinuity as homogeneous and unidirectional, $\init{\bm{v}}=-\init{u}\unit{z}$, which is strictly valid only if the bow shock is weak or absent.
The resulting discontinuity profile is approximately paraboloid at the nose, and quickly flattens to a $\myrho=2L_m$ cylinder far downstream.

The velocity field derived from Equation~(\ref{eq:RankineHalfSphere}) is
\begin{equation}
\bm{v} \equiv \begin{pmatrix} v_\myrho \\ v_\phi \\ v_z \end{pmatrix}
=
\init{u}\begin{pmatrix} \myrho \myPLS/r^3 \\ 0 \\ -1+z\myPLS/r^3 \end{pmatrix} \fin
\end{equation}
The stream functions may thus be chosen as $\lambda_1=\mygconst{R}\phi$ and
\begin{equation}
\lambda_2=\frac{1}{2\mygconst{R}}\left[\myrho^2+2\myPLS\left(\frac{z}{r}-1\right)\right]=\frac{\init{\myrho}^2}{2\mygconst{R}}\coma
\end{equation}
where, as above, the length scale $\mygconst{R}$ is arbitrary.
The thermal pressure is found from Bernoulli's Equation~(\ref{eq:Bernulli})
\begin{equation}
\myPT=\Prami\left(\Myichi+ \frac{L_m^2 z}{r^3}-\frac{L_m^4}{2r^4} \right)\coma
\end{equation}
where $\Pram\equiv \rho v^2$ is the ram pressure, and $\Mychi\equiv \myPT/\Pram\geq 3/2$ is necessary to guarantee a positive $\myPT$ throughout the flow.

The electric field outside the discontinuity is given, according to the axisymmetric flow Equation~(\ref{eq:E_Cylindrical}), by the simple analytic expression
  \begin{align}\label{eq:Esw}
    \bm{E}
= \frac{\myrho}{ \init{\myrho}c}\begin{pmatrix}[1.7]
              v_z\init{B}_\phi \\
             \frac{\init{\myrho}^2}{\myrho^2}\,  \init{u}\init{B}_\myrho \\
             - v_\myrho\init{B}_\phi
        \end{pmatrix}
        \fin
    \end{align}
The magnetic field can be written in terms of elliptic functions \citep{Roken2014}.
Close to the discontinuity $\init{\myrho}$ vanishes, so Equation~(\ref{eq:B_Cylindrical}) yields
      \begin{align}\label{eq:Bsw}
        \bm{B}
          &= \frac{L_m}{\init{\myrho}}\begin{pmatrix}[1.3]
                  -\init{B}_\myrho v_\myrho/\init{u}\\
                    \init{B}_\phi\myrho/L_m \\
                    -\init{B}_\myrho v_z/\init{u}
                \end{pmatrix} -\frac{\init{B}_z+\frac{3}{8}\init{B}_\myrho}{\init{u}}\begin{pmatrix}[1.3]
                              v_\myrho + O(\init{\myrho})\\
                               0 \\
                              v_z + O(\init{\myrho})
                          \end{pmatrix}
            \fin
        \end{align}

Near the discontinuity surface, where $\init{\myrho}$ is small, $\bm{E}$ is dominated by its $\myrho$ and $z$ components, and $\bm{B}$ is dominated by the first term in Equation~(\ref{eq:Bsw}).
In the cylindrical regime, $\bm{v}\to(-\init{u}\unit{z})$ and $\varrho\to2L_m$, so $E_\phi\ll E_\myrho\ll E_z$ and $B_\myrho\ll B_z\simeq (\init{B}_\myrho/2\init{B}_\phi)B_\phi$.

The magnetic field diverges near the discontinuity.
Hence, as in Section~\ref{sec:Boundarylayer}, we identify the region where $\epsilon_B>1$ as a highly-magnetized, self-limiting layer.
The outer edge of the magnetized layer, defined, say, by $\epsilon_B=1$, can now be used to estimate the electromagnetic fields in the discontinuity region $\Omega$, namely the heliopause or the magnetopause.

The width $\delta$ of the highly magnetized layer varies slightly along the discontinuity,
\begin{align}\label{eq:deltaT_MP}
\delta&=\left(\frac{2}{5+3\cos\angleBM}\right)^{1/2}\, f L_m
\coma
\end{align}
where $\angleBM$ is the polar angle, taken with respect to the $z$ axis with the object at the origin, and we defined
\begin{align}\label{eq:f_MP}
f\equiv\frac{\init{\epsilon}_B}{1 +\frac{1+4\cos\theta+3\cos2\theta}{16\Myichi}}
\fin
\end{align}
The electromagnetic fields on $\Omega$ are thus given by
\begin{align}\label{eq:E_alpha}
&\bm{E}_\Omega
= \frac{-\init{B}_\phi \init{u}}{c}\sqrt{\frac{f}{2+f}}\begin{pmatrix}[1.7]
              1+\frac{3-2\cos\theta-\cos2\theta}{f(2+f)}\\
            -\init{B}_\myrho/\init{B}_\phi \\
              \frac{4\cos^2(\theta/2)\sin\theta}{f(2+f)}
        \end{pmatrix}\\\nonumber
&=- \frac{2\init{B}_\phi \init{u}\sin^2(\theta/2)}{c\sqrt{f}}\begin{pmatrix}[1.7]
             1+\frac{\cos\theta}{2}\\
            0 \\
            \frac{\cos^3(\theta/2)}{\sin(\theta/2)}
        \end{pmatrix}+O\left(\sqrt{\init{\epsilon}_B}\right)\coma
\end{align}
and
\begin{equation}\label{eq:B_tinyEpB}
\bm{B}_\Omega= \frac{\init{B}_\myrho}{\sqrt{f}}\begin{pmatrix}[1.7]
           -\cos^3(\theta/2)\\
           -\init{B}_\phi/\init{B}_\myrho \\
           (1+\frac{\cos\theta}{2})\sin(\theta/2)
        \end{pmatrix}+O\left(\sqrt{\init{\epsilon}_B}\right)\fin
\end{equation}

As an example, consider the magnetopause of Earth.
Satellite measurements \citep[the 1999 $Wind$ spacecraft trajectory;][]{WangEtAl2003} suggest that halfway between the bow shock and the plasma depletion layer, $\init{u}\simeq300\km \se^{-1}$, $\init{B}\simeq0.2\,\mbox{mG}$, $\init{\bm{B}}\simeq(0.07,\,0.17,\,-0.05)\,\mbox{mG}$,
\begin{equation}
\frac{\MyPBi}{\Prami}=\frac{1}{2\init{M}_A^{2}}\simeq 0.1\coma
\end{equation}
and
\begin{equation}
\Myichi=\frac{\myPTi}{\Prami}=\frac{1}{\gamma\init{M}^2}\simeq 0.4\coma
\end{equation}
where $\MyPB\equiv B^2/(8\pi)$ is the magnetic pressure, and $M_A\equiv v/(B^2/4\pi\rho)^{1/2}$ is the Alfv$\acute{\mbox{e}}$n Mach number.
Hence the plasma is warm, and mildly magnetized with $\init{\epsilon}_B\simeq0.25$.

With these $\myH$ parameters, the passive evolution of the fields in the Rankine half-body model Equations~(\ref{eq:deltaT_MP})-(\ref{eq:B_tinyEpB}) yields \citep[at $\theta\sim 80^\circ$;][]{WangEtAl2003} $\delta\simeq 1.8\REarth$, $B_{\Omega}\simeq0.35\,\mbox{mG}$, and $E_{\Omega}\simeq10\,\mbox{mV/m}$.
These should be compared with the depletion layer thickness $\delta_{\mbox{\tiny dep}}\simeq 2.2 \REarth$ \citep{WangEtAl2003} and with the magnetopause fields.
Wind measured $B_\Omega=0.35\,\mbox{mG}$ \citep{WangEtAl2003}, but not the electric field; the $CLUSTER$ spacecraft measured $E_\Omega\simeq 10\,\mbox{mV/m}$ \citep[but three years layer, at $\theta\sim70^\circ$;][]{PanovEtAl2006A}.
The good agreement between measurements and model is largely coincidental, as the latter is crude, pertains only to the draped component, and fails to account for the complexity of the magnetosphere, the variability of the solar wind, and physical effects such as magnetic reconnection.

\section{Summary and Discussion}
\label{sec:Discussion}

We have shown that the stream functions $\lambda$ or path functions $\Lambda$, often used to describe non-diffusive flows, provide a simple, local (in $\lambda$ or $\Lambda$ space), and general solution to the problem of evolving electromagnetic fields in a highly conductive plasma, or equivalently propagating a length element $\bm{l}$, in an arbitrary flow.
Explicit expressions were provided, in both steady and time-dependent flow regimes, for the evolving magnetic field $\bm{B}$ (Equations~(\ref{eq:vecB_DeltaaT}) and (\ref{eq:B_patht})), electric field $\bm{E}$ (Equations~(\ref{eq:ESteady}) and (\ref{eq:Etimedependent})), and $\bm{l}$ (Equations~(\ref{eq:vecl_DeltaT}) and (\ref{eq:l_patht})).

The analysis, illustrated in Figure~\ref{fig:lambda_hat} for a steady flow, uses $\lambda$ or $\Lambda$ as coordinates.  Vectors are thus projected onto a reciprocal basis (Equations~(\ref{eq:a_basis}) or (\ref{eq:l_patht})), in which the vector components are conserved along streamlines or pathlines.
It is in this basis that the analysis becomes manifestly local, and thus conceptually and computationally simpler than previous solutions of the ideal MHD equations.
The results are more general than available until now because the closed-form expressions for $\bm{B}$ and $\bm{E}$ can be applied "as is", given $\Lambda$ or $\lambda$, for arbitrary magnetic field boundary conditions specified on any hypersurface $\myH$.
The analysis is valid for an arbitrary ideal MHD flow.

For steady flows, a third variable $q$ is needed, in addition to the two stream functions, in order to map the volume spanned by the flow.
Our analysis utilizes the freedom in the choice of this $q$; see Equations~(\ref{eq:vecl_Deltaq}) and (\ref{eq:vecB_Deltaq}).
Evolving $\bm{B}$ becomes considerably easier for certain choices of $q$, in particular if the travel time parameter $\Delta t$ is measured or easily computed; other choices require, in general, one integration (Equation~(\ref{eq:General_Deltat})).
Conveniently, the evolution of $\bm{E}$ is entirely fixed by the stream functions alone, given boundary conditions on $\myH$.

For time-dependent flows, the analysis is conceptually simpler because the three path functions entirely determine the flow.
The freedom in the choice of a fourth variable $Q$, needed to fully map spacetime, is less important. Here, no integration is necessary when $\Lambda_\mypathk$ are known.
The analysis is demonstrated for an arbitrary 1D flow (Equations~(\ref{eq:l_rarefaction}) and (\ref{eq:B_rarefaction})).

For both steady and time-dependent flows, computing the electric field is particularly simple in our method, as it only requires a differentiation of the stream or path functions.
This is much simpler than first computing the magnetic field, and then using Ohm's law to derive $\bm{E}$, because finding $\bm{B}$ typically involves a numerical integration (unless the path function are known).

The general expressions for the electromagnetic field allow us to quickly reproduce and generalize known magnetization solutions, such as for a potential incompressible flow around a sphere (Equations~(\ref{eq:B_incopote_vec}) and Figure~\ref{fig:PotIncSphere1}), and to derive new solutions for magnetization in various flows, such as an arbitrary axially symmetric flow (Equation~(\ref{eq:B_Cylindrical})), the incompressible Stokes flow around a sphere (Equations~(\ref{eq:B_Stokes}) and Figure~\ref{fig:stokesSphere1}), and the compressible flow in front of a blunt axisymmetric object (Equation~(\ref{eq:B_onaxis}) and Figure~\ref{fig:amplificaion_distance_all1}).

The magnetization in front of an axisymmetric blunt object (Equation~(\ref{eq:AxialB})) is of special interest in various astronomical systems, such as planetary bodies moving in a stellar wind, a stellar system moving through the ISM, or clumps and bubbles moving through the IGM.
The analysis simplifies considerably if the flow along the axis is approximately irrotational (as $C$ in Equation~(\ref{eq:AxialC}) vanishes).
The simple relation thus obtained between $\bm{B}$ and $v$ can be used to directly infer one quantity, if the other is known.
Here, the magnetic pressure (Equation~(\ref{eq:B_onaxis})) and magnetic pressure fraction (Equation~(\ref{eq:Bamplification})) increase toward the stagnation point, roughly as $v^{-1}$ (see Equation~(\ref{eq:StagnationB})), as illustrated for various flows in Figure~\ref{fig:amplificaion_distance_all1}.

Viscosity and compressibility are seen (in Figure~\ref{fig:amplificaion_distance_all1}) to enhance the magnetization, leading to a thicker magnetized layer.
In contrast, the magnetic pressure fraction appears to be quite insensitive to compressibility.
Crudely, the thickness of the magnetic layer depends on the upstream magnetization $\init{\epsilon}_B$ through Equation~(\ref{eq:ApproxBWidth}).

One advantage of our approach is that the results are directly generalized for nonhomogeneous electromagnetic field configurations upstream.
Indeed, given some spacetime-dependent $\init{\bm{B}}_{in}$ on $\myH$, the analysis remains intact, provided that $\init{\bm{B}}$ is determined by Equation~(\ref{eq:Binit_spacetime}).
This is illustrated in Figures~\ref{fig:PotIncSphereTSB} and \ref{fig:IPIPRL} for Fourier components of $\init{\bm{B}}_{in}$ in the stellar wind in front of a planetary body.
As a concrete example, we derive a semi-analytic expression (Equation~(\ref{eq:IPRL})) for the position of the IPRL behind a planetary body, and demonstrate (in Figure~\ref{fig:IPIPRL}) the substructure it acquires due to upstream magnetic variability.

Another advantage is that the electric field can be determined analytically in various flows.
As an illustration, we compute $\bm{E}$ (Equation (\ref{eq:Esw})) for a Rankine half-body model, approximately applicable above some heliospheres and magnetospheres.
Modeling the tangential discontinuity as a self-limiting magnetized layer, we estimate the draped electromagnetic component of the corresponding heliopause and magnetopause (Equations (\ref{eq:deltaT_MP})--(\ref{eq:B_tinyEpB})).

Our approach directly links the properties of electromagnetic fields to the properties of path functions and stream functions.
This connection, unexplored previously as far as we know, may thus help shed light on the general properties of both electromagnetic fields and path or stream functions, in complicated flow or field configurations.
Even if only one stream function is (or only 1--2 path functions are) known, our approach may still provide valuable constraints on the electromagnetic field evolution.
Vice versa, if the electromagnetic fields are known, the stream (or path) functions may be constrained, or even determined.

An important advantage of our analysis is its direct applicability to hydrodynamic numerical simulations that are already based on stream functions or path functions. Indeed, our results (in particular Equations~(\ref{eq:vecB_Deltaq}),(\ref{eq:ESteady}) or Equations~(\ref{eq:B_patht}),(\ref{eq:Etimedependent})) can be immediately incorporated into such simulations, in order to follow the passive evolution of electromagnetic fields. Moreover, our results suggest that the back-reaction of electromagnetic fields on the flow could be easily added to such a code, thus leading to an efficient MHD simulation.

Consider, for example, a putative path function simulation for a 3D time-dependent flow.
It is most natural to choose the hypersurface $\myH$ as the 3D volume at some initial time; this corresponds to imposing initial conditions on the magnetic field, in addition to the initial conditions on the flow.
As the grid coordinates correspond to the path functions, the 3D magnetic configuration is simply frozen onto the grid.
Thus, the magnetic field does not need to be evolved, and its back reaction on the flow can be easily computed.
Implementation schemes are deferred to future work.

Hydrodynamic simulations based on stream functions or path functions have been slowly evolving over the past few decades.
Stream function codes have been developed for steady flows in 3D \citep[\eg][]{Pearson1981,ChungyuanDulikravich1986,SherifHafez1988,Beale1993,KnightMallinson1996,ElshabkaChung1999}.
Path function simulations have been so far presented for time-dependent flows in 1D \citep{Hui2007} and in 2D \citep[][]{LohHui2000}.

Like any Lagrangian code, such simulations have the advantage of precisely conserving the mass flux, matching the resolution to the evolving flow, and sharply resolving discontinuities.
Employing path functions has the additional advantage of converting the hydrodynamic equations into a precise, geometric, conservative form.
This bears considerable numerical advantages, and guarantees the correct deformation of each grid cell, although strong cell distortion \eg in vortical flow, must still be addressed.
This has been explicitly shown in 2D \citep{LohHui2000}, and partly developed for 3D \citep{Van-RoesselHui1991}.
An MHD simulation based on path function will thus have important advantages in both hydrodynamic and electromagnetic sectors, and would supplement present MHD codes.

\acknowledgements
We thank Yuri Lyubarski and Michael Gedalin for helpful discussions.
This research was funded by the European Union Seventh Framework
Programme (FP7/2007-2013) under grant agreement
n\textordmasculine~293975, an IAEC-UPBC joint research foundation grant, and an ISF-UGC grant.



\appendix

\newcommand{\myrelation}{\bm{B}\propto \rho \bm{l}}

\section{The relation $\protect\myrelation$ preserves the MHD equations}\label{sec:app_Gauss}

We evolve the electromagnetic fields using Equation~(\ref{eq:generalB}), or equivalently the Helmholtz equation (\ref{eq:Helmholtz}), so it is useful to prove that the resulting $\bm{B}$ satisfies Gauss' law (\ref{eq:Gauss}) and the convection equation (\ref{eq:convection}).
We assume that the Helmholtz equation and the continuity equation (\ref{eq:continuity}) are satisfied everywhere, and that Gauss' law holds on $\myH$.

First note that
\begin{equation}\label{eq:convection_not_validated}
\frac{\partial\boldsymbol{B}}{\partial t}-\grad\times\left(\boldsymbol{v}\times\boldsymbol{B}\right)
=\frac{\partial\boldsymbol{B}}{\partial t}
-\left(\boldsymbol{B}\cdot\grad\right)\boldsymbol{v}
+\left(\boldsymbol{v}\cdot\grad\right)\boldsymbol{B}
+\boldsymbol{B}\left(\grad\cdot\boldsymbol{v}\right)
-\boldsymbol{v}\left(\grad\cdot\boldsymbol{B}\right)
=(-\boldsymbol{v})\left(\grad\cdot\boldsymbol{B}\right)\coma
\end{equation}
where we used the Helmholtz and continuity equations in the last step.
This ensures that if Gauss' law is satisfied, then the convection equation (\ref{eq:convection}) is guaranteed, and vice versa.
Taking the divergence of this equation yields
\begin{equation}\label{eq:dgauss_dt}
\frac{d\left(\grad\cdot\boldsymbol{B}\right)}{dt}=-\left(\grad\cdot\boldsymbol{v}\right)\left(\grad\cdot\boldsymbol{B}\right)\fin
\end{equation}
With the initial conditions $\grad\cdot\bm{B}=0$ assumed on $\myH$, the trivial solution Equation~(\ref{eq:Gauss}) spans the entire flow connected to $\myH$.
Gauss law is thus satisfied, and, according to Equation~(\ref{eq:convection_not_validated}), so is the convection equation.

\section{The integral $(\partial_\lambda\deltats)_\myr$ for a steady incompressible potential flow around a sphere }
\label{sec:AppPerpBPotIncFlow}

The integral in Equation~(\ref{eq:F_of_incopote}) can be solved by expanding the integrand about $\lambda=0$.
This expansion is useful for analyzing the flow in the close vicinity of the axis of symmetry or of the sphere, where $\lambda$ is small.
The result is
\begin{align}
& \int\frac{\myr^4d\myr}{\left(\myr^3-1\right)^2\left(1-\frac{2\lambda\myr}{\myr^3-1} \right)^{\frac{3}{2}}} \,
= \sum_{n=0}^{\infty} b_n(\myr) (2\lambda)^{n} \coma
\end{align}
where $n$ is an integer,
\begin{equation}
b_n(\myr)  =  \frac{2\,\Gamma\left(n+\frac{3}{2}\right)(-1)^{n}\myr^{n+2}}{\,n!(n+2)\sqrt{\pi}}
 \left[
\mbox{}_2F_1\left(\frac{2+n}{3},2+n;\frac{5+n}{3};\myr^3\right)-\mbox{}_2F_1\left(1+n,\frac{2+n}{3};\frac{5+n}{3};\myr^3\right)
\right] \coma
\end{equation}
$\Gamma$ is the Euler Gamma function, and $\mbox{}_2F_1$ is the hypergeometric function. In particular,
\begin{equation}
b_0=\frac{1}{9}\left\{2\sqrt{3}\arctan\left(\frac{1+2\myr}{\sqrt{3}}\right)+\log\left[\frac{\left(\myr-1\right)^2}{1+\myr+\myr^2}\right]-\frac{3\myr^2}{\myr^3-1}\right\},
\qquad \mbox{and} \qquad
b_1=\frac{1-2\myr^3}{4\left(\myr^3-1\right)^2} \fin
\end{equation}


\end{document}

%% file: Definitions.tex
\newcommand{\ie}{\emph{i.e.} }
\newcommand{\eg}{\emph{e.g.,} }
\newcommand{\cf}{\emph{cf.} }

\newcommand{\be}{\begin{equation}}
\newcommand{\ee}{\end{equation}}
\newcommand{\bea}{\begin{equation*}}
\newcommand{\eea}{\end{equation*}}
\newcommand{\beqr}{\begin{eqnarray} \nonumber}
\newcommand{\eeqr}{\end{eqnarray}}
\newcommand{\beqrb}{\begin{eqnarray}}
\newcommand{\eeqrb}{\nonumber \end{eqnarray}}
\newcommand{\fin}{\mbox{ .}}
\newcommand{\coma}{\mbox{ ,}}

\newcommand{\se}{\mbox{ s}}

\newcommand{\km}{\mbox{ km}}

\newcommand{\initial}[1]{\tilde{#1}}

\newcommand{\grad}{\bm{\nabla}}

\newcommand{\unit}[1]{\bm{\hat{#1}}}

\newcommand{\myr}{{r}}

\newcommand{\myj}{{j}}